\documentclass[pra,aps,preprint,amsmath,amssymb,showpacs]{revtex4}
\usepackage{graphicx}
\usepackage{epsfig}
\usepackage{amsbsy}
\include{graphics}

\begin{document}

\title{Deterministic Raman crosstalk effects in amplified 
wavelength division multiplexing transmission}

\author{Quan M. Nguyen and Avner Peleg}

\affiliation{Department of Mathematics, State University of New York
at Buffalo, Buffalo, New York 14260, USA}


\begin{abstract}
We study the deterministic effects of Raman-induced crosstalk 
in amplified wavelength division multiplexing (WDM)  optical fiber 
transmission lines. We show that the dynamics of pulse amplitudes 
in an $N$-channel transmission system is described by an 
$N$-dimensional predator-prey model. We find the equilibrium 
states with non-zero amplitudes and prove their  
stability by obtaining the Lyapunov function. The stability is 
independent of the exact details of the approximation for the 
Raman gain curve. Furthermore, we investigate the impact of cross 
phase modulation and Raman self and cross frequency shifts on the 
dynamics and establish the stability of the equilibrium state with 
respect to these perturbations. Our results provide a quantitative 
explanation for the robustness of differential-phase-shift-keyed WDM 
transmission against Raman crosstalk effects. 
\end{abstract}


\maketitle 

\section{Introduction}
One of the important nonlinear processes affecting pulse propagation 
in massive wavelength division multiplexing (WDM) optical fiber 
communication systems is due to inter-pulse Raman-induced crosstalk 
\cite{Agrawal2001,Stolen80}. In this process, which takes place 
during collisions between pulses from different frequency channels 
(interchannel collisions), energy is transferred from 
high-frequency pulses to low-frequency pulses. 
It is known that the magnitude of the Raman-induced energy exchange in 
a single interchannel collision is independent of the frequency difference 
between the channels. Consequently, the magnitude of the cumulative 
energy shifts for a given pulse grows with the square of the 
number of channels, a result that is valid for linear transmission 
\cite{Chraplyvy84,Tkach95,Jander96,Ho2000}, conventional soliton 
transmission \cite{Chi89,Malomed91a,Kumar98,P2004}, and strongly 
dispersion-managed (DM) soliton transmission \cite{Kaup99}. 
Therefore, in a 100-channel system, for example, the Raman crosstalk 
effects can be larger by a factor of $2.5\times 10^3$ compared with a 
two-channel system operating at the same bit rate per channel.

Early studies of Raman crosstalk in WDM transmission focused on the 
dependence of the energy shifts on the total number of channels 
\cite{Chraplyvy84}, as well as on the impact of energy 
depletion \cite{Jander96} and group velocity dispersion 
\cite{Cotter84,Sarkar2007} on the dynamics. 
Later on attention turned to the combined effects of Raman crosstalk 
and bit-pattern randomness in the on-off-keying (OOK) transmission scheme,  
and it was found that the probability density function (PDF) of the 
pulse amplitudes is lognormal 
\cite{Tkach95,Ho2000,Kumar2003,P2004,CP2005,Yamamoto2003}. 
This finding means that the $n$th normalized moments of the amplitude 
grow exponentially with both propagation distance and $n^{2}$. 
Furthermore, in several studies of conventional soliton transmission 
it was found that the dynamics of the frequency shift is 
strongly coupled to amplitude dynamics, and as a result, the $n$th 
normalized moments of the Raman-induced self and cross frequency shifts 
also grow exponentially with propagation distance and $n^{2}$ 
(see Refs. \cite{P2007,CP2008,P2009}). 
This intermittent dynamic behavior has important 
practical consequences, by leading to relatively high 
bit-error-rate (BER) values at intermediate and large propagation 
distances \cite{P2007,CP2008,P2009}. 
Raman crosstalk effects were also recently investigated 
in hut-skipped amplified WDM transmission \cite{Zhou2006,Mazroa2009}, 
in cable television overlay passive optical networks \cite{Tian2006,Kim2007}, 
in optical code-division multiple-access transmission 
systems \cite{Mokhtar2009}, and in conjunction with 
four-wave-mixing \cite{Vanholsbeeck2005}.

One of the ways to overcome the detrimental effects of Raman crosstalk 
on massive WDM transmission is by replacing the OOK scheme by alternative 
encoding schemes, which are expected to be less susceptible to these effects. 
The differential phase shift keying (DPSK) scheme, in which all time slots 
are occupied and the information is encoded in the phase difference 
between adjacent pulses, is one of the promising encoding methods 
that have attracted much interest in recent years \cite{Xu2004,Gnauck2005}.  
In DPSK transmission the Raman-induced amplitude dynamics becomes 
(approximately) deterministic, and important questions arise regarding the 
character of this dynamics. One major question concerns the 
possibility to achieve a stable equilibrium state for the amplitudes in 
all channels. The study reported in Ref. \cite{Jander96} demonstrated that 
this is not possible in unamplified optical fiber lines. However, later 
experiments showed that the situation is quite different in amplified 
WDM transmission \cite{Golovchenko2000,Kim2008}. 
More specifically, it was found that the introduction 
of amplification into the system significantly 
reduces the Raman-induced energy shifts. In the present paper we suggest 
a dynamical explanation for this important experimental observation. 
Moreover, we demonstrate the robustness of DPSK transmission against 
inter-pulse Raman crosstalk effects by showing that equilibrium states 
with non-zero amplitudes in all channels do exist, and by proving 
the stability of the equilibrium states.

In the present study we consider optical solitons as an example for 
the pulses carrying the information for the following reasons. 
First, as mentioned above, the Raman-induced energy exchange in 
pulse collisions is similar in linear transmission, conventional 
soliton transmission, and strongly DM soliton transmission. Second, 
propagation of conventional solitons through an optical fiber is 
described by the nonlinear Schr\"odinger (NLS) equation, which is an 
integrable model \cite{Zakharov72}. Due to the integrability of the model, 
and to the fact that optical solitons are stable stationary solutions 
of the NLS equation, the derivation of the model for the 
Raman-induced amplitude dynamics can be done in a rigorous manner. 
Third, conventional optical solitons have traditionally been considered 
as excellent candidates for information transmission in high-speed 
optical fiber lines and in all-optical networks \cite{Agrawal2001}. 
Furthermore, state-of-the-art transmission experiments already use 
all-Raman distributed amplification \cite{Islam2004,Agrawal2005,
Mollenauer2003,Grosz2004,Mamyshev2004,Rottwitt2004,Turitsyn2006}, 
which is the most suitable amplification scheme for conventional solitons.

Since we consider transmission systems where the pulses in each 
frequency channel are well-separated, intrachannel 
four-wave-mixing (FWM) effects are negligible. In addition, we assume 
that loss is compensated by distributed Raman amplification. 
It is a well-known fact in soliton theory that in the absence of loss 
interchannel FWM products completely vanish after the collisions, 
see Ref. \cite{Zakharov72} for theory and Refs. \cite{Chi89,MM98} 
for numerical simulations with the NLS equation with and 
without delayed Raman response. We remark that even in systems 
employing lumped amplifiers (and non-overlapping pulses), 
interchannel FWM can be significantly reduced by dispersion-management. 
This is true both in the soliton regime \cite{MM98} and in the linear 
regime \cite{Gnauck2008}. In contrast, the energy 
shift in a single collision between two strongly DM  
solitons is given by an expression with exactly the same 
form as the expression obtained for conventional solitons \cite{Kaup99}, 
i.e., strong dispersion-management does not reduce Raman crosstalk.

In deriving the model for Raman-induced amplitude dynamics we fully 
take into account pulse walk-off. In addition, we assume 
that the pulse sequences in all frequency channels are deterministic and 
that the sequences are either infinitely long or are subject to periodic 
temporal boundary conditions. The first setup approximates long-haul 
transmission, while the second one corresponds to a closed fiber loop 
experiment. We also assume that the constant net gain/loss in each 
channel is determined by the difference between distributed amplifier 
gain and fiber loss. Notice that in this feature our model is 
fundamentally different from the model derived in Ref \cite{Jander96}. 
Indeed, since in the latter model all channels experience net loss, 
it does not support an equilibrium state with non-zero amplitudes in 
all channels. In contrast, in our model the net gain/loss of some  
channels is positive while for other channels it is negative, 
which is the underlying reason for the existence of equilibrium 
states with non-zero amplitudes in all channels.

Our model for the Raman crosstalk dynamics in an $N$-channel system 
consists of a system of $N$ coupled nonlinear ordinary differential 
equations (ODEs) for the amplitudes in different channels. 
The system of coupled ODEs can be described in the 
jargon of population dynamics theory as an $N$-dimensional 
predator-prey model \cite{Volterra26}.   
After obtaining the model we look for equilibrium states with 
non-zero amplitudes in all channels and establish their stability 
with respect to deviations of the initial amplitudes from the  
equilibrium values. We also investigate the dynamic behavior 
induced by such deviations. In actual optical fiber transmission systems 
pulse dynamics can be influenced by physical processes other than 
Raman crosstalk. It is therefore important to understand the manner in 
which these additional processes perturb the Raman-induced amplitude 
dynamics. In the present study we investigate the effects of three 
perturbations due to cross phase modulation (XPM), Raman self frequency 
shift (SFS), and Raman cross frequency shift (XFS). For each of these 
physical processes we construct the perturbed model describing 
amplitude dynamics in an $N$-channel system and investigate the 
existence of equilibrium states with non-zero amplitudes in all channels. 
Furthermore, we study the stability of the equilibrium state 
and the typical dynamic behavior of the amplitudes in two-channel systems 
in the presence of each of the three perturbations.

The rest of the paper is organized as follows. 
In Section \ref{derivation} we derive the unperturbed model for 
Raman-induced amplitude dynamics in WDM systems with $2N+1$ channels, 
and in Section \ref{stability} we find 
the equilibrium states of the model and investigate their stability.
In Section \ref{simulations} we validate the predictions of 
Section \ref{stability} by numerical simulations.    
The perturbed models with XPM, Raman SFS, and Raman XFS, are studied 
in Sections \ref{XPM}, \ref{SFS}, and \ref{XFS}, respectively, for 
two-channel systems. Section \ref{conclusions} is reserved for conclusions. 
In Appendix \ref{appendB} we obtain the perturbed models with XPM, 
Raman SFS, and Raman XFS in WDM transmission with $2N+1$ channels.

\section{Dynamics of deterministic Raman crosstalk - unperturbed model}
\label{unperturbed}
\subsection{Derivation of the model}
\label{derivation} 
Propagation of short pulses of light through an optical fiber 
in the presence of delayed Raman response is described by 
the following perturbed NLS equation \cite{Agrawal2001}:
\begin{eqnarray}
i\partial_z\psi+\partial_t^2\psi+2|\psi|^2\psi=
-\epsilon_{R}\psi\partial_t|\psi|^{2},
\label{cfs1}
\end{eqnarray}
where $\psi$ is proportional to the envelope of the 
electric field, $z$ is propagation distance and $t$ 
is time in the retarded reference frame. 
The term $-\epsilon_{R}\psi\partial_t|\psi|^{2}$ represents 
the first order approximation for the fiber's delayed Raman 
response \cite{Stolen89} and $\epsilon_{R}$ is the 
Raman coefficient \cite{dimensions}.
When $\epsilon_{R}=0$, the single-soliton solution of 
Eq. (\ref{cfs1}) in a frequency channel $\beta$ is given by
\begin{eqnarray}
\psi_{\beta}(t,z)\!=\!
\eta_{\beta}\frac{\exp(i\chi_{\beta})}{\cosh(x_{\beta})},
\label{cfs2}
\end{eqnarray}
where $x_{\beta}=\eta_{\beta}\left(t-y_{\beta}-2\beta z\right)$, 
$\chi_{\beta}=\alpha_{\beta}+\beta(t-y_{\beta})+
\left(\eta_{\beta}^2-\beta^{2}\right)z$, 
and $\eta_{\beta}, \alpha_{\beta}$ and $y_{\beta}$ are the 
soliton amplitude, phase and position, respectively.

Let us describe the main Raman-induced effects on a single collision 
between a soliton from the $j$th frequency channel and a 
soliton from the $k$th frequency channel. We assume that 
$\epsilon_{R}\ll 1/|\beta_{j}-\beta_{k}| \lesssim 1$, 
which is the typical situation in many WDM transmission systems even for 
adjacent channels \cite{MM98,Gnauck2008}. Under this assumption one can 
show that the most important effect of delayed Raman response on the 
collision is an $O(\epsilon_{R})$ change in the soliton amplitude 
\cite{Chi89,Malomed91a,Kumar98,P2004,CP2005} 
\begin{eqnarray}
\Delta\eta_{j}=
2\epsilon_{R}f(|j-k|)\mbox{sgn}(\beta_{k}-\beta_{j})\eta_{j}\eta_{k},
\label{crosstalk3}
\end{eqnarray}
where the coupling constants $f(|j-k|)$ depend on the 
specific approximation for the Raman gain curve. 
Notice that due to the inclusion of the $f(|j-k|)$ factors, 
Eq. (\ref{crosstalk3}) does not rely on the triangular 
approximation for the Raman gain curve. If we adopt the triangular 
approximation, we obtain that $f(|j-k|)=1$ for any $j$ and $k$. 
The  effects of the collision in order 
$\epsilon_{R}/|\beta_{j}-\beta_{k}|$ will be described in 
Section \ref{perturbed}, where we obtain the perturbed 
models for amplitude dynamics. 
Since $\epsilon_{R}\ll 1/|\beta_{j}-\beta_{k}| \lesssim 1$, effects of 
order $\epsilon_{R}^{2}$ and higher will be neglected.   
In addition, third order dispersion and self-steepening are neglected 
since the collision-induced effects of these conservative perturbations 
on the soliton amplitude and frequency are of higher order in both the 
parameter $\epsilon$ characterizing the perturbative process 
and $1/|\beta|$ (see, e.g., Refs. \cite{Malomed91b,PCG2003,PCG2004}).

Consider now WDM transmission systems with $2N+1$ channels and frequency 
difference $\Delta\beta$ between adjacent channels. Our model, which 
takes into account pulse walk-off, is based on the following assumptions. 
(1) The soliton sequences in all channels are deterministic in the sense 
that all time slots are occupied and each soliton is located at the 
center of a time slot of width $T$. Furthermore, when considering amplitude 
dynamics, the amplitudes are equal for all pulses from the same frequency 
channel, but are not necessarily equal for pulses from different channels. 
This setup corresponds, for example, to return-to-zero (RZ) 
transmission with differential-phase-shift-keying \cite{OOK}. 
(2) The sequences are either (a) infinitely long, or (b) subject to 
periodic temporal boundary conditions. Notice that setup (a) 
is an approximation for long-haul transmission systems, 
while setup (b) is an approximation for closed fiber-loop experiments. 
(3) The gain/loss in each channel is determined 
by the difference between distributed amplifier gain and fiber loss. 
In particular, for some channels this difference can be slightly positive, 
resulting in small net gain, while for other channels this difference can be 
slightly negative, resulting in small net loss. We emphasize that in 
this feature our model is essentially different from the models 
that are usually considered in studies of Raman amplification schemes, 
where it is assumed that {\it all} pumps experience net loss 
\cite{Agrawal2001,Jander96}.

To obtain the dynamic equation for the amplitude of the 
$j$th-channel solitons we note that the distance traveled by 
these solitons while passing two successive solitons in the $j-1$ 
or $j+1$ channels is $\Delta z_{c}^{(1)}=T/(2\Delta\beta)$. 
We denote by $z_{l}$ the location of the $l$th collision of a given 
$j$th-channel soliton with solitons in the $j+1$ or $j-1$ channel. 
Using Eq. (\ref{crosstalk3}) and summing over all collisions 
occurring within the interval $(z_{l-1},z_{l}]$, where 
$z_{l}=z_{l-1}+\Delta z_{c}^{(1)}$, we obtain    
\begin{eqnarray} &&
\!\!\!\!\!\!\!\!\!\!\eta_{j}(z_{l-1}+\Delta z_{c}^{(1)})=
\eta_{j}(z_{l-1})+g_{j}\eta_{j}(z_{l-1})\Delta z_{c}^{(1)}+ 
2\epsilon_{R}\!\!\!\sum_{k=-N}^{N}
(k-j)f(|j-k|)\eta_{j}(z_{l-1})\eta_{k}(z_{l-1}). 
\label{crosstalk4}
\end{eqnarray}     
The constant $g_{j}$ on the right hand side of Eq. (\ref{crosstalk4}) 
is the net gain/loss coefficient for the $j$th channel, which is 
assumed to be independent of $z$. Due to the periodicity of the 
pulse sequences the same equation is satisfied by the amplitudes 
of all $j$th-channel solitons. Moreover, the same equations with 
different $j$ values, where $j=-N, \dots, N$, describe the dynamics 
of the soliton amplitudes in all channels. 
Going to the continuum limit we obtain       
\begin{eqnarray} &&
\frac{d \eta_{j}}{dz}=
\eta_{j}\left[g_{j}+ 
C\sum_{k=-N}^{N}(k-j)f(|j-k|)\eta_{k}\right],  
\label{crosstalk5}
\end{eqnarray}   
where $C=4\epsilon_{R}\Delta\beta/T$ and $j=-N, \dots, N$. 
The system (\ref{crosstalk5}) gives a 
complete description of the dynamics of the $\eta_{j}$'s. Notice that 
apart from the important fact that some gain/loss coefficients can be 
positive while others can be negative, the system (\ref{crosstalk5}) 
is similar to the one obtained in Ref. \cite{Jander96} for the 
Raman-induced amplitude dynamics of continuous waves in unamplified 
WDM transmission. Thus, our model also describes the Raman-induced 
crosstalk dynamics of continuous waves in amplified 
WDM transmission systems.

\subsection{Equilibrium states, stability, and conserved quantities}
\label{stability} 
In optical fiber communication systems it is usually desired to achieve 
a steady state in which the pulse amplitudes in all channels 
are equal and constant (independent of $z$) \cite{Agrawal2001}. 
We therefore look for an equilibrium state of the system (\ref{crosstalk5}) 
in the form $\eta^{(eq)}_{j}=\eta>0$ for $-N \le j \le N$. 
Setting the right hand sides of (\ref{crosstalk5}) equal to zero 
we arrive at  
\begin{eqnarray} &&
g_{j}=-C\eta\sum_{k=-N}^{N}(k-j)f(|j-k|).  
\label{crosstalk6}
\end{eqnarray}      
Therefore, the gain required to maintain an equilibrium state 
with equal amplitudes is {\it not} flat with respect to frequency. 
Instead, the net gain/loss coefficient of high-frequency channels should 
be positive, while that of low-frequency channels should be negative. 
In other words, high-frequency channels should be overamplified, 
whereas low-frequency channels should be underamplified compared with the 
reference ($j=0$) channel. 
Substituting Eq. (\ref{crosstalk6}) into Eq. (\ref{crosstalk5}) 
we arrive at a slightly simpler form of the model, which is convenient 
for analysis and numerical simulations,  
\begin{eqnarray} &&
\frac{d \eta_{j}}{dz}=
C\eta_{j}\sum_{k=-N}^{N}(k-j)f(|j-k|)(\eta_{k}-\eta).  
\label{crosstalk7}
\end{eqnarray}   
Equation (\ref{crosstalk7}) can be described in population dynamics 
terminology as a predator-prey system with $2N+1$ species \cite{Volterra26}.

The equilibrium states of Eq. (\ref{crosstalk7}) with non-zero amplitudes 
are determined by 
\begin{eqnarray} &&
\sum_{k=-N}^{N}(k-j)f(|j-k|)(\eta_{k}^{(eq)}-\eta)=0, 
\;\;\;\;\;\; -N \le j \le N.  
\label{crosstalk7a}
\end{eqnarray}   
The trivial solution of Eq. (\ref{crosstalk7a}), that is, the solution 
with $\eta^{(eq)}_{k}=\eta>0$ for $-N \le k \le N$, corresponds to the 
equilibrium state of Eq. (\ref{crosstalk7}) with equal non-zero amplitudes. 
However, due to the anti-symmetry of $(k-j)f(|j-k|)$ with respect to 
an interchange of $j$ and $k$, Eq. (\ref{crosstalk7a}) has infinitely many 
non-trivial solutions, and these correspond to equilibrium states of 
Eq. (\ref{crosstalk7}) with unequal non-zero amplitudes \cite{2N}. 
In the case where the Raman gain curve is described by the 
triangular approximation it is straightforward to show that 
the non-trivial equilibrium states of Eq. (\ref{crosstalk7}) 
are determined by the following two equations: 
\begin{eqnarray} &&
\sum_{k=-N}^{N} \eta_{k}^{(eq)}=(2N+1)\eta,
\;\;\;\;\;\;\;\sum_{k=-N}^{N} k\eta_{k}^{(eq)}=0. 
\label{crosstalk7b}
\end{eqnarray}              
Therefore, in this case the equilibrium states lie on a 
$(2N-1)$-dimensional plane. Thus, for a 3-channel system, 
for example, the equilibrium states lie on the line segment 
$(\eta,\eta,\eta)-(b-\eta)(1/2,-1,1/2)$, where $0<b<3\eta$.

We now prove stability of all equilibrium states of 
Eq. (\ref{crosstalk7}) with non-zero amplitudes,    
$\eta_{j}=\eta_{j}^{(eq)}>0, -N \le j \le N$,  
by constructing a Lyapunov function for this equation.  
For this purpose we consider the function   
\begin{eqnarray} &&
V_{L}(\pmb{\eta})=\sum_{j=-N}^{N}
\left[\eta_{j}-\eta_{j}^{(eq)}
+\eta_{j}^{(eq)}\ln\left(\frac{\eta_{j}^{(eq)}}{\eta_{j}}\right)\right],
\label{crosstalk8}
\end{eqnarray}      
where $\pmb{\eta}=(\eta_{-N},\dots,\eta_{j}, \dots, \eta_{N})$.
Taking the derivative of $V_{L}(\pmb{\eta})$ along trajectories 
of the system (\ref{crosstalk7}) we obtain  
\begin{eqnarray} &&
\frac{d V_{L}}{dz}=
\sum_{j=-N}^{N}\frac{\eta_{j}-\eta_{j}^{(eq)}}{\eta_{j}}
\frac{d \eta_{j}}{dz}=
C\sum_{j=-N}^{N}(\eta_{j}-\eta_{j}^{(eq)})
\sum_{k=-N}^{N}(k-j)f(|j-k|)(\eta_{k}-\eta).
\label{crosstalk9}
\end{eqnarray}    
From Eq. (\ref{crosstalk7a}) it follows that 
$\sum_{k=-N}^{N}(k-j)f(|j-k|)(\eta_{k}-\eta)=
\sum_{k=-N}^{N}(k-j)f(|j-k|)(\eta_{k}-\eta^{(eq)})$. 
Using this relation together with Eq. (\ref{crosstalk9}) 
and the anti-symmetry of $(k-j)f(|j-k|)$ we arrive at 
\begin{eqnarray} &&
\frac{d V_{L}}{dz}=
\sum_{j=-N}^{N}\sum_{k=-N}^{N}(k-j)f(|j-k|)
(\eta_{j}-\eta_{j}^{(eq)})(\eta_{k}-\eta^{(eq)})=0,  
\label{crosstalk10}
\end{eqnarray}    
for $\eta_{j}>0, -N \le j \le N$. It is straightforward to show that 
each term $h(\eta_{j})=\eta_{j}-\eta_{j}^{(eq)}+
\eta_{j}^{(eq)}\left[\ln(\eta_{j}^{(eq)}/\eta_{j})\right]$ 
on the right hand side of Eq. (\ref{crosstalk8}) 
satisfies $h(\eta_{j})\ge 0$ for any $\eta_{j}>0$, where 
$h(\eta_{j})=0$ only at $\eta_{j}=\eta_{j}^{(eq)}$. 
Therefore, $V_{L}(\pmb{\eta})\ge 0$ 
for any vector $\pmb{\eta}$ for which $\eta_{j}>0$ for $-N \le j \le N$,
where equality holds only at equilibrium points. Combining this result  
with the result $dV_{L}/dz=0$ along trajectories of the system we 
conclude that $V_{L}$ is a Lyapunov function of the system 
(\ref{crosstalk7}) and therefore the equilibrium states 
$\eta_{j}=\eta^{(eq)}_{j}>0, -N \le j \le N$ 
are stable \cite{Smale74,Wiggins91}.
We note that since $dV_{L}/dz=0$ rather than $dV_{L}/dz<0$ this 
stability means that the values of $\eta_{j}(z)$ are bounded for 
any $z$ but do not tend to $\eta_{j}^{(eq)}$ for large propagation 
distance $z$. Thus, the typical dynamics of the amplitudes $\eta_{j}(z)$ 
for initial conditions that are off the equilibrium point is oscillatory.     
Notice that the stability is independent of the exact specification of 
the $f(|j-k|)$ values, and therefore the equilibrium states are stable 
irrespective of the specific details of the approximation for the 
Raman gain curve. Notice also that Eq. (\ref{crosstalk8}) actually 
provides $K+1$ independent conserved quantities for the model 
(\ref{crosstalk7}), where $K$ is the dimension of the solution space 
of Eq. (\ref{crosstalk7a}). For example, if the Raman gain is described 
by the triangular approximation, $K=2N-1$, and there are $2N$ conserved 
quantities in total.

\subsection{Numerical solution of the system (\ref{crosstalk7})}
\label{simulations}
In order to check the predictions of the previous subsection we 
solve Eq. (\ref{crosstalk7}) numerically by employing a 
fourth-order Runge-Kutta method. As a concrete example for the physical 
parameter values we consider a 101-channel system operating 
at 40 Gbits/s per channel with frequency spacing $\Delta\nu=100$ GHz
and dimensionless time slot width $T=5$. With this choice 
the pulse width is 5 ps, $\Delta\beta=\pi$, $\epsilon_{R}=0.0012$, 
and $N=50$. These parameter values are typical 
for several state-of-the-art massive multichannel transmission experiments, 
see Ref. \cite{Gnauck2008} and references therein. 
Taking $\beta_{2}=-1\mbox{ps}^{2}\mbox{km}^{-1}$ 
and $\gamma=4\mbox{W}^{-1}\mbox{km}^{-1}$ we obtain $P_{0}=10$ mW for 
the soliton peak power.  
The dimensionless final propagation distance is taken as $z_{f}=200$ 
corresponding to $X_{f}=10^{4}$ km, but the main features of the 
dynamics can already be observed at considerably shorter distances.         
We also choose $\eta=1$, so that the trivial equilibrium state is 
$\eta_{j}^{(eq)}=1$, for all $j$.

\begin{figure}[ptb]
\begin{tabular}{cc}
\epsfxsize=8.0cm  \epsffile{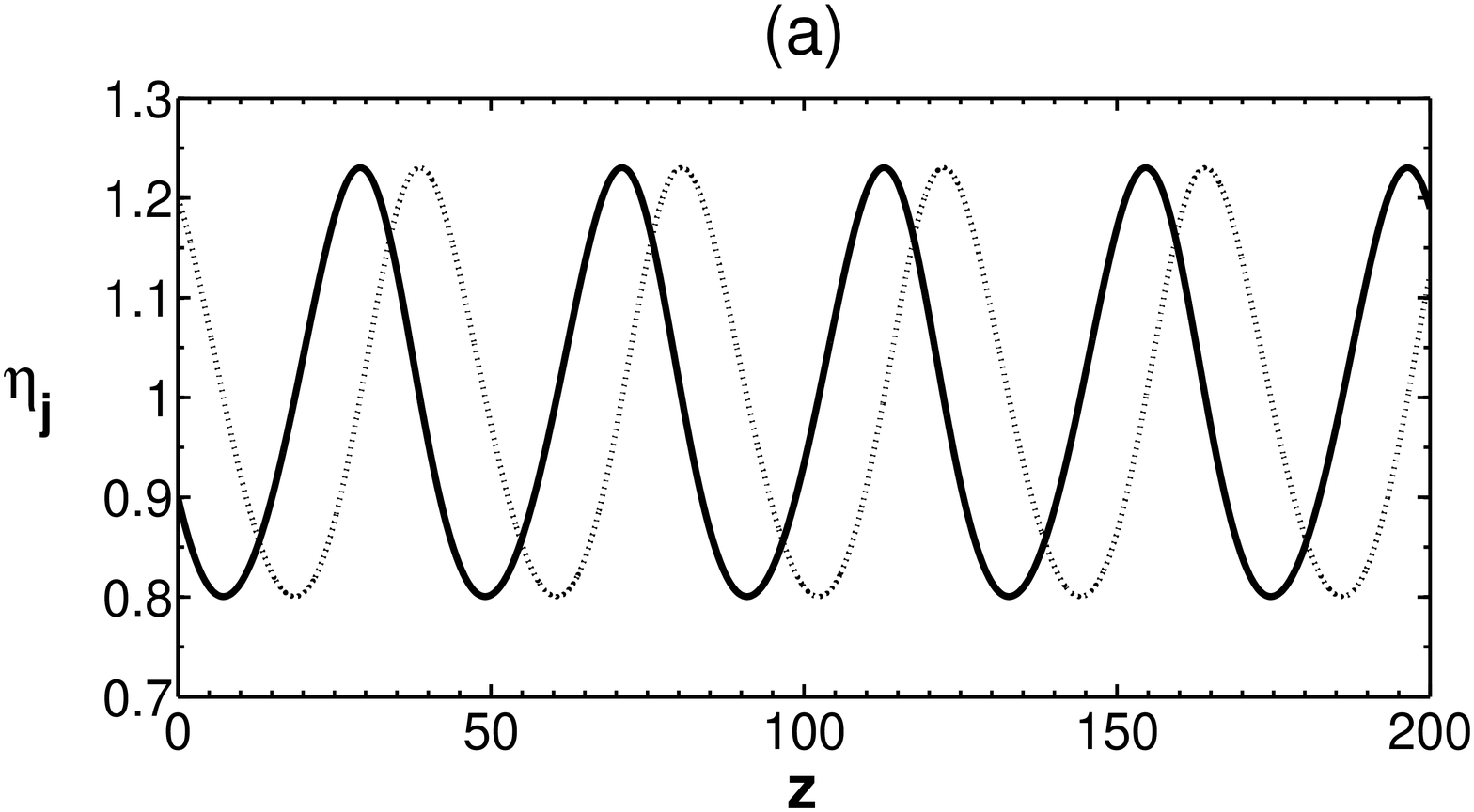} &
\epsfxsize=8.0cm  \epsffile{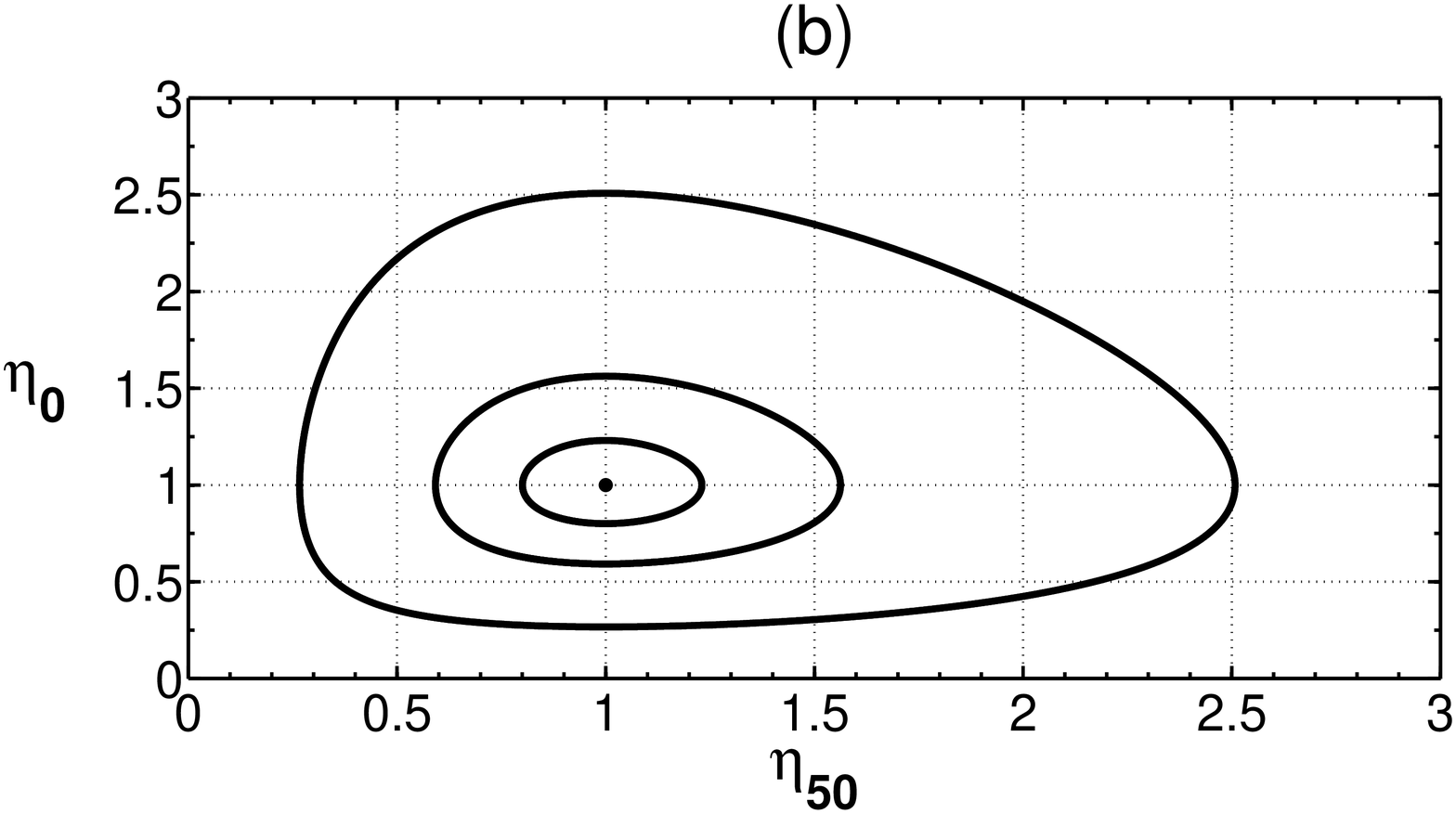} 
\end{tabular}
\caption{Amplitude dynamics for the two-channel system described by 
Eq. (\ref{simu1}). (a) The $z$-dependence of $\eta_{50}$ (solid curve) 
and $\eta_{0}$ (dotted curve) with the initial condition 
$\eta_{50}(0)=1.2$ and $\eta_{0}(0)=0.9$. 
(b) The corresponding phase portrait.}
 \label{fig1}
\end{figure}

To illustrate the impact of Raman crosstalk on {\it massive} 
WDM transmission we focus attention on amplitude dynamics of solitons 
from faraway channels. Thus, we start by considering a two-channel system 
consisting of the reference channel ($j=0$) and the highest-frequency 
channel $j=N=50$. Employing Eq. (\ref{crosstalk7}) to this two-channel 
system while adopting the triangular approximation for the Raman gain 
curve we obtain 
\begin{eqnarray} &&
\frac{d \eta_{50}}{dz}=
50C\eta_{50}\left(1-\eta_{0}\right),
\nonumber \\&&
\frac{d \eta_{0}}{dz}=
50C\eta_{0}\left(\eta_{50}-1\right).  
\label{simu1}
\end{eqnarray}   
The solution of Eq. (\ref{simu1}) for $\eta_{0}>0$ and $\eta_{1}>0$ 
is 
\begin{eqnarray} &&
\eta_{50}(z)+\eta_{0}(z)-\ln\left[\eta_{50}(z)\right]-
\ln\left[\eta_{0}(z)\right]=\kappa,
\label{simu1a}
\end{eqnarray}  
where the constant $\kappa$ is determined by the values of 
$\eta_{0}(0)$ and $\eta_{50}(0)$.  
The $z$-dependence of the amplitudes $\eta_{0}$ and $\eta_{50}$ is 
shown in Fig. \ref{fig1} (a)  for the initial condition $\eta_{50}(0)=1.2$ 
and $\eta_{0}(0)=0.9$ . It is clearly seen that the amplitudes 
oscillate about their equilibrium value $\eta=1$. The dimensionless 
oscillation period is $Z_{p}=41.8$, corresponding to 2090 km \cite{Z_p}. 
Similar oscillatory behavior is observed for other choices of 
the initial amplitudes. Figure \ref{fig1} (b) shows the phase 
portrait for the system $(\ref{simu1})$. Since all trajectories 
are closed curves we conclude that the dynamics is indeed periodic 
and that the equilibrium state is stable.

\begin{figure}[ptb]
\begin{tabular}{cc}
\epsfxsize=8.0cm  \epsffile{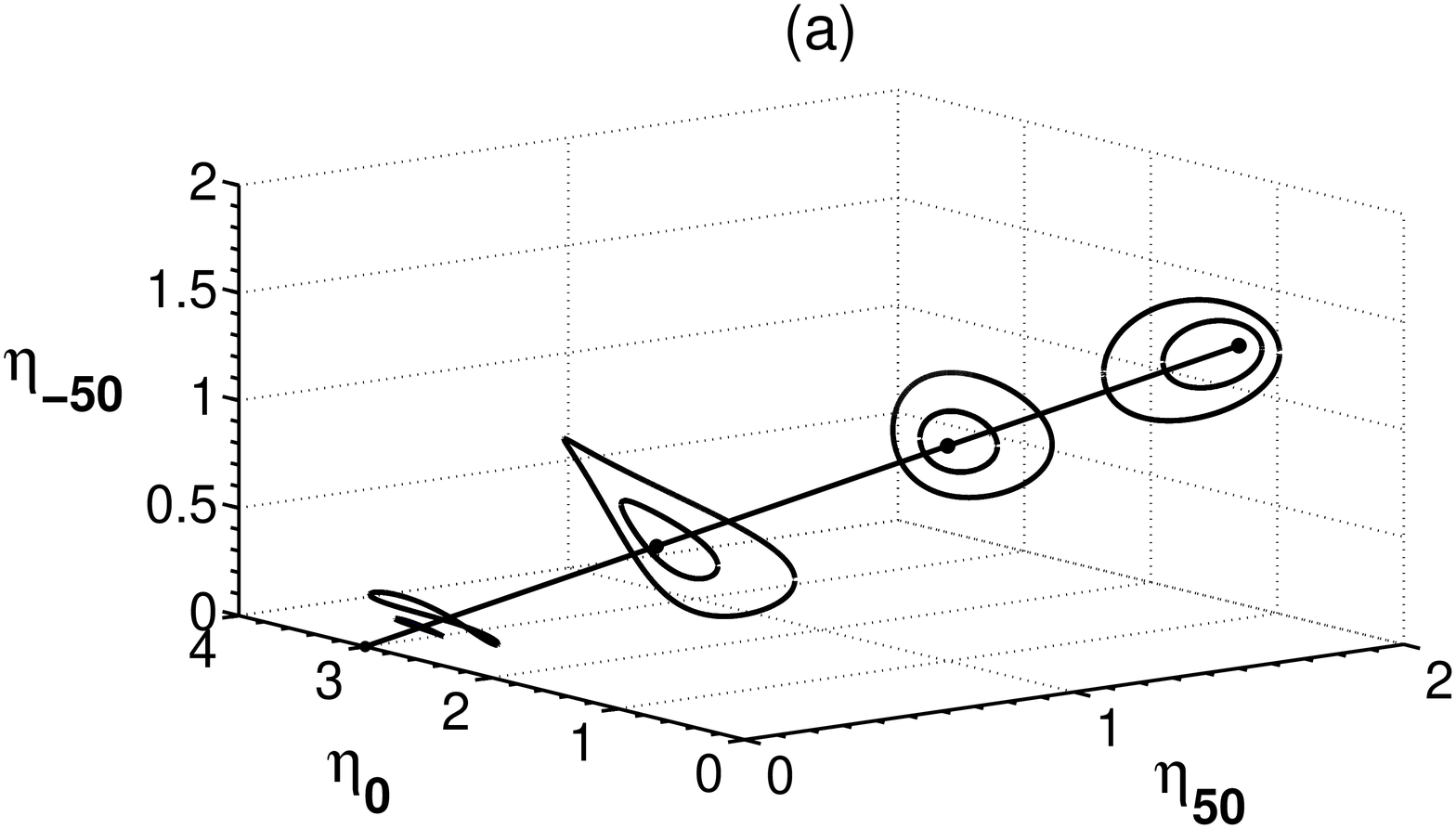} &
\epsfxsize=8.0cm  \epsffile{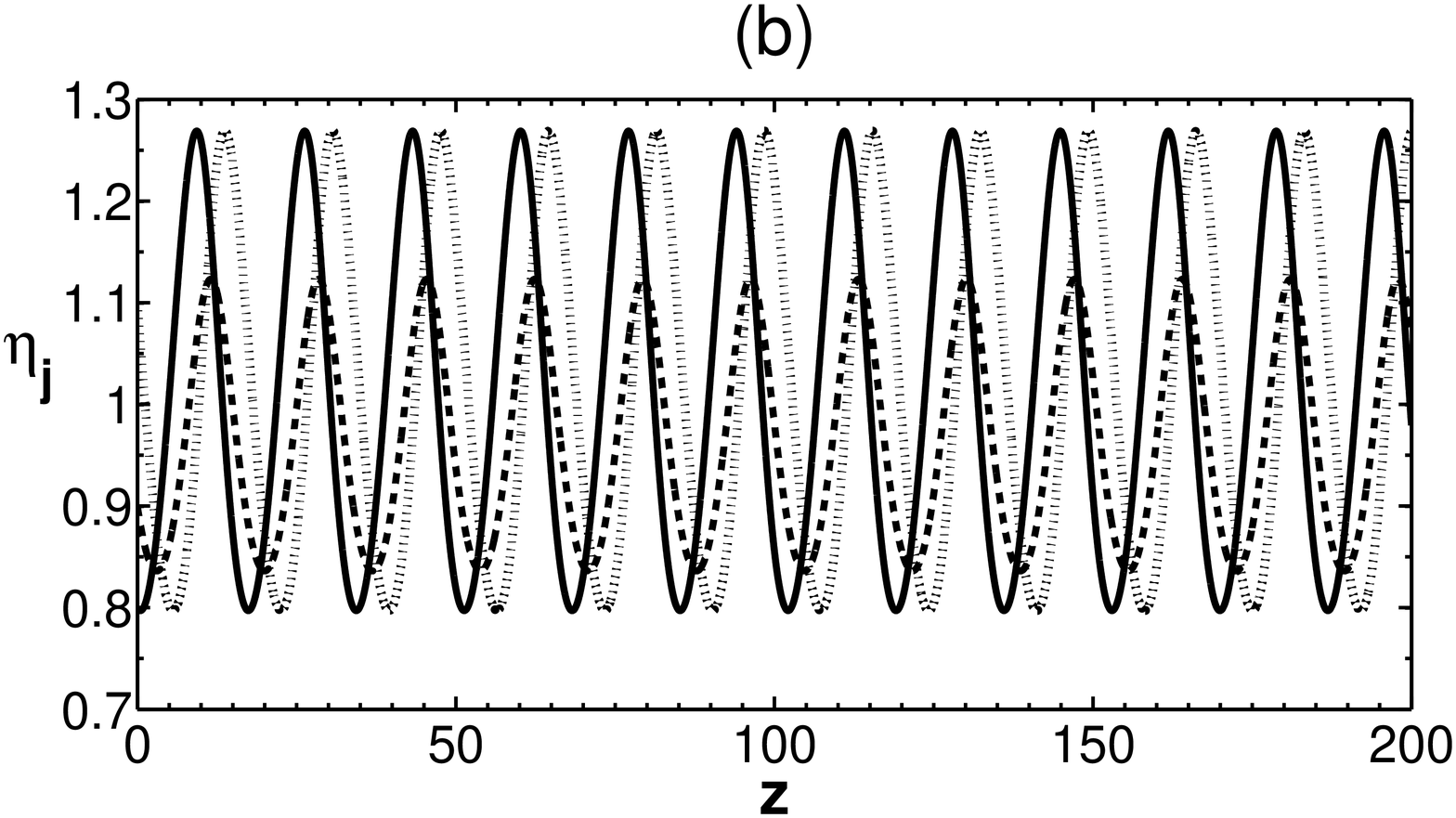} 
\end{tabular}
\caption{Amplitude dynamics for the three-channel system described by 
Eq. (\ref{simu2}). (a) The phase portrait. (1,1,1) is the second 
equilibrium point from the right. 
(b) The $z$-dependence of $\eta_{50}$ (solid curve), 
$\eta_{0}$ (dotted curve), and $\eta_{-50}$ (dashed curve), with the 
initial condition  $\eta_{50}(0)=0.8$, $\eta_{0}(0)=0.9$, 
$\eta_{-50}(0)=1.1$. }
 \label{fig3}
\end{figure}

Next we consider a three-channel system consisting of the channels 
$j=0$ and $j=\pm 50$. Within the Raman triangular approximation 
the amplitude dynamics is described by 
\begin{eqnarray} &&
\frac{d \eta_{50}}{dz}=
50C\eta_{50}\left(3-\eta_{0}-2\eta_{-50}\right),
\nonumber \\&&
\frac{d \eta_{0}}{dz}=
50C\eta_{0}\left(\eta_{50}-\eta_{-50}\right),
\nonumber \\&&
\frac{d \eta_{-50}}{dz}=
50C\eta_{-50}\left(-3+2\eta_{50}+\eta_{0}\right).  
\label{simu2}
\end{eqnarray}  
Figure \ref{fig3} (a) shows the phase portrait. 
The solid line is the line segment of stable equilibrium points 
$(1,1,1)-(b-1)(1/2,-1,1/2)$, where $0<b<3$.   
Since all trajectories are closed orbits centered about the 
equilibrium line we conclude that the equilibrium states are 
indeed stable and that the amplitudes exhibit periodic oscillations. 
This oscillatory dynamics is illustrated in Fig. \ref{fig3} (b) 
for the initial condition $\eta_{50}(0)=0.8$, $\eta_{0}(0)=0.9$, 
$\eta_{-50}(0)=1.1$.

As another example for possible dynamic scenarios exhibited by the 
unperturbed model (\ref{crosstalk7}) we study a four-channel 
system consisting of the channels $j=\pm 16$ and $j=\pm 48$. 
As we demonstrate below, a four-channel setup represents the 
simplest case where deviations from the triangular approximation 
for the Raman gain curve lead to the emergence of 
new dynamical features. We consider the following sets of values 
for the coupling constants $f(|j-k|)$: $f(1)=f(2)=1$ and $f(3)=1+p$, 
where $p=0$ for set (1) and $p=0.1$ for set (2). Thus, set (1) 
corresponds to the triangular approximation, whereas set (2) represents 
a 10$\%$ deviation from the triangular approximation for $f(3)$.      
For these values of the coupling constants the dynamics of the 
amplitude is described by 
\begin{eqnarray} &&
\frac{d \eta_{48}}{dz}=
32C\eta_{48}\left[(6+3p)-\eta_{16}-2\eta_{-16}-3(1+p)\eta_{-48}\right],
\nonumber \\&&
\frac{d \eta_{16}}{dz}=
32C\eta_{16}\left(2+\eta_{48}-\eta_{-16}-2\eta_{-48}\right),
\nonumber \\&&
\frac{d \eta_{-16}}{dz}=
32C\eta_{-16}\left(-2+2\eta_{48}+\eta_{16}-\eta_{-48}\right),
\nonumber \\&&
\frac{d \eta_{-48}}{dz}=
32C\eta_{-48}\left[-(6+3p)+3(1+p)\eta_{48}+2\eta_{16}+\eta_{-16}\right].  
\label{simu7}
\end{eqnarray}   
We solve the system (\ref{simu7}) numerically with $p=0$ and $p=0.1$ 
and with the initial condition $\eta_{48}(0)=1.2$, $\eta_{16}(0)=1.1$, 
$\eta_{-16}(0)=0.95$, and $\eta_{-48}(0)=0.9$. The results of our 
numerical simulations are presented in Figs. \ref{fig5} and \ref{fig6}. 
As can be seen, in both cases the amplitudes exhibit oscillatory 
dynamics. However, in the triangular approximation case ($p=0$) there 
is only a single period for the oscillations, whereas in the non-triangular 
approximation case ($p=0.1$) two very different oscillation periods are 
observed. The first period is not significantly different from the 
oscillation period for $p=0$, while the second period is much longer than 
the first one (see Fig. \ref{fig6}). In fact, in order to verify that 
the amplitude dynamics is indeed periodic for $p=0.1$ one has to carry out  
the simulation up to a final propagation distance of $z_{f}=2000$.

Notice that our choice of 40 Gbits/s per channel systems is mainly for 
reasons of convenience, since with this value we can illustrate the 
oscillatory nature of the Raman-induced amplitude dynamics already with 
two and three channels. Since current soliton-based systems work at 10 
Gbits/s per channel it is useful to examine one example for the 
amplitude dynamics in such systems. We therefore consider 
a 361-channel system operating at 10 Gbits/s per channel with 
frequency spacing $\Delta\nu=25$ GHz and dimensionless time slot 
width $T=5$. For these parameter values the pulse width is 20 ps, 
$\Delta\beta=\pi$, $\epsilon_{R}=3\times 10^{-4}$, and $N=180$.
Taking $\beta_{2}=-2\mbox{ps}^{2}\mbox{km}^{-1}$ and 
$\gamma=2\mbox{W}^{-1}\mbox{km}^{-1}$ we obtain $P_{0}=2.5$ mW for 
the soliton peak power. The dimensionless final propagation distance is 
taken as $z_{f}=25$ corresponding to $X_{f}=10^{4}$ km, but the main 
dynamical features can already be observed at shorter distances.         
As before we choose $\eta=1$, so that the trivial equilibrium state is 
$\eta_{j}^{(eq)}=1$, for all $j$. To illustrate the dynamics we 
consider a seven-channel system consisting of the channels 
$j=0$, $j=\pm 60$, $j=\pm 120$, and $j=\pm 180$ and adopt the 
triangular approximation for the Raman gain curve.   
The corresponding system of equations for amplitude evolution 
is solved numerically with the initial condition 
$\eta_{180}(0)=1.2$, $\eta_{120}(0)=1.05$, $\eta_{60}(0)=1.1$ 
$\eta_{0}(0)=1.15$, $\eta_{-60}(0)=0.98$, $\eta_{-120}(0)=1.1$, 
and $\eta_{-180}(0)=0.95$. The results are shown in 
Fig. \ref{7channels}. One can see that pulse amplitudes in all 
channels exhibit oscillatory behavior similar to the one observed 
in Figures \ref{fig1} and \ref{fig3}  for the 40 
Gbits/s per channel system. Thus, equilibrium states with non-zero 
amplitudes in all channels are stable in both 10 and 40 Gbits/s per 
channel systems, in agreement with the predictions 
in Section \ref{stability}.

\section{Dynamics of deterministic Raman crosstalk - perturbed models}
\label{perturbed}
We now turn to study the effects of perturbations on the model 
described by Eq. (\ref{crosstalk7}). We focus attention on the effects 
of cross phase modulation (XPM), Raman self frequency shift (SFS), 
and  Raman cross frequency shift (XFS). It is well-known that 
each of these processes by itself does not change the soliton amplitude. 
Thus, in the absence of Raman crosstalk the soliton amplitudes are 
constant and there is no amplitude dynamics to be considered. 
As we show below XPM, Raman SFS, and Raman XFS affect amplitude 
dynamics by leading to $z$-dependence of the collision rate. 
For these reasons, the three processes can be considered as 
perturbations to the model (\ref{crosstalk7}).
For each of the three processes we develop a perturbed model for 
amplitude dynamics and analyze the stability of the equilibrium state 
and the dynamic behavior in the new model. 
Here we choose to concentrate on a two-channel system 
consisting of channels $j=0$ and $j=1$ since such treatment is 
sufficient in order to uncover the main changes in the dynamics 
compared with the unperturbed model. Without loss of generality we 
assume $f(1)=1$ for the coupling constant. The general forms of 
the perturbed models in WDM systems with $2N+1$ channels are obtained 
in Appendix \ref{appendB}.

\subsection{Effects of cross phase modulation}
\label{XPM}
The XPM-induced position shift experienced by a soliton in the $j$th 
channel as a result of a single collision with a $k$th-channel soliton 
is given by (see, e.g., Refs. \cite{MM98,CP2005}) 
\begin{eqnarray} &&
\Delta y_{j}=\frac{4\mbox{sgn}(\beta_{k}-\beta_{j})\eta_{k}}
{(\beta_{k}-\beta_{j})^{2}}.
\label{crosstalk26}
\end{eqnarray}     
Our goal is to obtain a perturbed model for the dynamics of the amplitudes 
$\eta_{0}$ and $\eta_{1}$ in the two-channel system, which takes into 
account  the XPM-induced position shift.  
It is clear that the main effect of the position shift (\ref{crosstalk26}) 
is to lead to a change in the inter-collision distance $\Delta z_{c}^{(1)}$, 
that is, $\Delta z_{c}^{(1)}$ is $z$-dependent in the perturbed model. 
In order to find this $z$-dependence we first write down the equation 
for the location of the $l$th collision of the soliton from 
the zero time slot in the reference channel. 
Since this collision is with the soliton from the $-l$ time slot 
in channel 1, the collision distance $z_{l}$ is determined by 
\begin{eqnarray}
y_{1,-l}(z_{l})=y_{0,0}(z_{l}),
\label{crosstalk27}
\end{eqnarray} 
where the first subscript in $y_{1,-l}$ stands for the channel and 
the second subscript represents the time slot. Taking into account 
the difference in group velocities, $2\beta_{1}$, and the position 
shifts experienced by the solitons in the collision we find 
\begin{eqnarray}
y_{1,-l}(z_{l-1})+2\beta_{1}\Delta z_{c}^{(1)}(z_{l-1})-
4\eta_{0}(z_{l-1})/\beta_{1}^{2}=y_{0,0}(z_{l-1})
+4\eta_{1}(z_{l-1})/\beta_{1}^{2}.
\label{crosstalk28}
\end{eqnarray} 
Solving Eq. (\ref{crosstalk28}) for $\Delta z_{c}^{(1)}$ 
we obtain     
\begin{eqnarray}
\Delta z_{c}^{(1)}(z_{l-1})=
\frac{1}{2\beta_{1}}\left[y_{0,0}(z_{l-1})-y_{1,-l}(z_{l-1})\right]
+\frac{2}{\beta_{1}^{3}}
\left[\eta_{0}(z_{l-1})+\eta_{1}(z_{l-1})\right].
\label{crosstalk29}
\end{eqnarray}
From the definition of $z_{l-1}$ and the periodicity of the 
pulse sequences it follows that $y_{0,0}(z_{l-1})-y_{1,-l}(z_{l-1})=T$. 
Therefore, Eq. (\ref{crosstalk29}) can be rewritten as
\begin{eqnarray}
\Delta z_{c}^{(1)}(z_{l-1})=\frac{T}{2\beta_{1}}
\left\{1+B\left[\eta_{0}(z_{l-1})+\eta_{1}(z_{l-1})\right]\right\},
\label{crosstalk30}
\end{eqnarray} 
where $B=4/(T\beta_{1}^{2})$.
Notice that $\Delta z_{c}^{(1)}(z_{l-1})> T/(2\beta_{1})$ due to the 
fact that the XPM-induced position shifts are positive for the 
reference channel solitons and negative for the solitons in channel 1.  
The equations for the amplitudes $\eta_{0}$ and $\eta_{1}$ 
at $z_{l}=z_{l-1}+\Delta z_{c}^{(1)}(z_{l-1})$ are similar in form 
to Eq. (\ref{crosstalk4}). Substituting relation (\ref{crosstalk30}) 
into the equations for $\eta_{0}(z_{l-1}+\Delta z_{c}^{(1)})$ 
and $\eta_{1}(z_{l-1}+\Delta z_{c}^{(1)})$ and going to 
the continuum limit we obtain 
\begin{eqnarray} &&
\frac{d \eta_{1}}{dz}=
\eta_{1}\left[g_{1}-\frac{C\eta_{0}}
{1+B\left(\eta_{0}+\eta_{1}\right)}\right],
\nonumber \\&&
\frac{d \eta_{0}}{dz}=
\eta_{0}\left[g_{0}+\frac{C\eta_{1}}
{1+B\left(\eta_{0}+\eta_{1}\right)}\right].
\label{crosstalk31}
\end{eqnarray}        
Equation (\ref{crosstalk31}) represents the perturbed model that 
takes into account XPM effects. Notice that for the typical values 
$T=5$ and $\beta_{1}=\pi$, $B\simeq 0.081$, i.e., $B\ll 1$. 
Thus, for such values the XPM perturbation can be considered as a weak 
perturbation.     

We look for equilibrium states of Eq. (\ref{crosstalk31}) that are of 
the form $\eta^{(eq)}_{0}=\eta^{(eq)}_{1}=\eta>0$. This requirement 
yields the expressions $g_{1}=-g_{0}=C\eta/(1+2B\eta)$ for the gain/loss 
coefficients. Thus, an important consequence of the XPM-induced position 
shifts is a change in the values of the gain/loss coefficients that are 
required for maintaining an equilibrium state with equal non-zero amplitudes. 
Taking into account the modified expressions for $g_{0}$ and $g_{1}$,  
we can rewrite Eq. (\ref{crosstalk31}) as    
\begin{eqnarray} &&
\frac{d \eta_{1}}{dz}=
C\eta_{1}\left[\frac{\eta}{1+2B\eta}-\frac{\eta_{0}}
{1+B\left(\eta_{0}+\eta_{1}\right)}\right],
\nonumber \\&&
\frac{d \eta_{0}}{dz}=
C\eta_{0}\left[-\frac{\eta}{1+2B\eta}
+\frac{\eta_{1}}
{1+B\left(\eta_{0}+\eta_{1}\right)}\right].
\label{crosstalk32}
\end{eqnarray}             
To prove stability of the equilibrium point $(\eta,\eta)$ we look 
for a Lyapunov function of the system (\ref{crosstalk32}) in the 
form 
\begin{eqnarray} &&
V_{L}(\eta_{0},\eta_{1})=
\eta\ln\left(\frac{\eta}{\eta_{0}}\right)+
\eta\ln\left(\frac{\eta}{\eta_{1}}\right)+
\frac{1+2B\eta}{B}
\ln\left[\frac{1+B(\eta_{0}+\eta_{1})}{1+2B\eta}\right].
\label{crosstalk33}
\end{eqnarray}      
Taking the derivative along trajectories of the system (\ref{crosstalk32}) 
we arrive at 
\begin{eqnarray} &&
\frac{d V_{L}}{dz}=
-\frac{\eta}{\eta_{0}}\frac{d \eta_{0}}{dz}
-\frac{\eta}{\eta_{1}}\frac{d \eta_{1}}{dz}
+\frac{1+2B\eta}{1+B(\eta_{0}+\eta_{1})}
\frac{d }{dz}(\eta_{0}+\eta_{1}).
\label{crosstalk34}
\end{eqnarray}    
Replacing $d\eta_{0}/dz$ and $d\eta_{1}/dz$ with the right hand sides 
of Eq. (\ref{crosstalk32}), we find  
\begin{eqnarray} &&
\frac{d V_{L}}{dz}=
-C\eta\left[-\frac{\eta}{1+2B\eta}+\frac{\eta_{1}}
{1+B\left(\eta_{0}+\eta_{1}\right)}\right]
-C\eta\left[\frac{\eta}{1+2B\eta}-\frac{\eta_{0}}
{1+B\left(\eta_{0}+\eta_{1}\right)}\right]
\nonumber \\&&
-\frac{C\eta(\eta_{0}-\eta_{1})}{1+B\left(\eta_{0}+\eta_{1}\right)}=0.
\label{crosstalk34a}
\end{eqnarray}       
Hence, $V_{L}$ is constant along trajectories of Eq. (\ref{crosstalk32}). 
In addition, it is straightforward to show that $V_{L}$ attains its 
minimum at $(\eta,\eta)$. Since $V_{L}(\eta,\eta)=0$, it follows that  
$V_{L}(\eta_{0},\eta_{1})\ge 0$ for any $(\eta_{0},\eta_{1})$ such 
that $\eta_{0}>0$ and $\eta_{1}>0$. 
Combining these results we conclude that $V_{L}$ is a Lyapunov function 
for Eq. (\ref{crosstalk32}) and that the equilibrium state $(\eta,\eta)$ 
is a center \cite{Smale74,Wiggins91}. Consequently, $(\eta,\eta)$ is 
a stable equilibrium point and deviations of the initial amplitude values  
from $\eta$ lead to oscillatory dynamics of $\eta_{0}(z)$ and $\eta_{1}(z)$. 
This means that the XPM perturbation does not change the 
stability properties of the equilibrium state. 

To illustrate these conclusions we numerically solve Eq. 
(\ref{crosstalk32}) by use of a fourth-order Runge-Kutta scheme. 
For concreteness we consider a 2-channel transmission system operating 
at 160 Gbits/s per channel with time slot width $T=5$. In this system  
the pulse width is 1.25 ps and $\epsilon_{R}=0.0048$. Notice  
that WDM transmission at 160 Gbits/s per channel has  
received much attention in recent years both theoretically \cite{Cuenot2003} 
and experimentally \cite{Gnauck2003,Morita2006,Sandel2007}. 
Taking $\beta_{2}=-0.5\mbox{ps}^{2}\mbox{km}^{-1}$ and 
$\gamma=4\mbox{W}^{-1}\mbox{km}^{-1}$ we obtain $P_{0}=80$ mW for 
the soliton peak power. The dimensionless final propagation 
distance is taken as $z_{f}=1280$ corresponding to propagation 
over $X_{f}=8000$ km, but the main dynamical features are observed 
already at much shorter distances. Without loss of generality we 
choose $\eta=1$, so that the equilibrium state is $(1,1)$.  
Figure \ref{fig7} shows the $z$-dependence 
of $\eta_{1}$ and $\eta_{0}$ with the initial condition 
$\eta_{1}(0)=1.1$ and $\eta_{0}(0)=0.9$ for $\Delta\beta=2.0$ (a) 
and $\Delta\beta=10.0$ (b). The latter values of $\Delta\beta$ 
correspond to interchannel frequency spacing of $\Delta\nu=250$ GHz 
and $\Delta\nu=1250$ GHz, respectively. In both cases the amplitudes 
exhibit oscillations about the equilibrium value in accordance with 
the prediction of the analytic calculations. Similar oscillatory 
dynamics is observed for other initial conditions and other values of 
$\Delta\beta$. Furthermore, as can be seen from Fig. \ref{fig8},  
all solutions curves in the phase plane are closed. Thus, our 
numerical simulations validate the predictions about stability 
of the equilibrium state in the presence of XPM.

\subsection{Effects of Raman self frequency shift}
\label{SFS}
The Raman-induced SFS experienced by the $j$th-channel 
solitons is given by \cite{Mitschke86,Gordon86,Kodama87} 
\begin{eqnarray}&&
\frac{d\beta_{j}}{dz}=-\frac{8}{15}\epsilon_{R}\eta_{j}^{4}(z),
\label{crosstalk35}
\end{eqnarray}
where we explicitly take into account the coupling of the frequency shift 
to the amplitude. Consider the effects of the Raman SFS on amplitude 
dynamics in a two-channel system consisting of channels 0 and 1.
Since in general $\eta_{0}(z) \ne \eta_{1}(z)$, 
the Raman-induced SFS can lead to $z$-dependence of the frequency 
difference $\beta_{10}(z)=\beta_{1}(z)-\beta_{0}(z)$, which in turn 
would lead to $z$-dependence of the inter-collision distance 
$\Delta z_{c}^{(1)}=T/(2\beta_{10})$. 
Therefore, the impact of the Raman SFS on amplitude dynamics in a 
two-channel system can be taken into account by replacing the constant 
frequency difference $\Delta\beta$ by the $z$-dependent frequency difference 
$\beta_{10}(z)$ in the equations for $\eta_{0}$ and $\eta_{1}$,  
and by using relation (\ref{crosstalk35}) to obtain the dynamic 
equation for $\beta_{10}$. This calculation yields the following system:  
\begin{eqnarray} &&
\frac{d \eta_{1}}{dz}=
\eta_{1}\left(g_{1}-\frac{C}{\Delta\beta}\beta_{10}\eta_{0}\right),
\nonumber \\&&
\frac{d \eta_{0}}{dz}=
\eta_{0}\left(g_{0}+\frac{C}{\Delta\beta}\beta_{10}\eta_{1}\right),
\nonumber \\&&
\frac{d\beta_{10}}{dz}=-\frac{8}{15}\epsilon_{R}
\left(\eta_{1}^{4}-\eta_{0}^{4}\right).
\label{crosstalk36}
\end{eqnarray}      
Looking for equilibrium states of the system (\ref{crosstalk36}) 
in the form $(\eta, \eta, \Delta\beta)$ we obtain $g_{1}=-g_{0}=C\eta$ for 
the gain/loss coefficients. Comparing this result with the result obtained 
in section \ref{unperturbed} for the unperturbed model we see that the 
Raman-induced SFS does not change the values of the gain/loss coefficients 
that are required for maintaining an equilibrium state with 
equal amplitudes. Using the values $g_{1}=-g_{0}=C\eta$ we can 
rewrite the system (\ref{crosstalk36}) in a simpler form,  
\begin{eqnarray} &&
\frac{d \eta_{1}}{dz}=
C\eta_{1}\left(\eta-\beta_{10}\eta_{0}/\Delta\beta\right),
\nonumber \\&&
\frac{d \eta_{0}}{dz}=
C\eta_{0}\left(-\eta+\beta_{10}\eta_{1}/\Delta\beta\right),
\nonumber \\&&
\frac{d\beta_{10}}{dz}=-\frac{8}{15}\epsilon_{R}
\left(\eta_{1}^{4}-\eta_{0}^{4}\right).
\label{crosstalk37}
\end{eqnarray}     
We study the stability of the equilibrium state $(\eta, \eta, \Delta\beta)$ 
of (\ref{crosstalk37}) by linear stability analysis and by numerical 
simulations. Linear stability analysis predicts a bifurcation at 
$\Delta\beta_{bif}=(16T\eta^{3}/15)^{1/2}$. 
For $\Delta\beta>\Delta\beta_{bif}$ the Jacobian matrix of 
(\ref{crosstalk37}) has two purely imaginary eigenvalues and one zero 
eigenvalue, while for $\Delta\beta<\Delta\beta_{bif}$ 
all three eigenvalues are real. In the latter case one eigenvalue is positive, 
another is negative, and the third one is zero. 
For $\Delta\beta=\Delta\beta_{bif}$ all three eigenvalues 
are equal to zero. Based on this analysis one might suspect that the 
equilibrium state $(\eta, \eta, \Delta\beta)$ becomes unstable for 
$\Delta\beta<\Delta\beta_{bif}$. However, since $(\eta, \eta, \Delta\beta)$ 
remains a non-hyperbolic equilibrium point, linear stability 
calculations might fail \cite{Smale74,Wiggins91} and one has to 
resort to numerical simulations to study stability. 

We perform numerical simulations with Eq. (\ref{crosstalk37}) for the 
same two-channel system that was considered in subsection \ref{XPM}. 
Choosing $\eta=1$, the equilibrium state is $(1, 1, \Delta\beta)$ and 
the bifurcation value predicted by linear stability analysis 
is $\Delta\beta_{bif}\simeq 2.31$. Figure \ref{fig9} shows the 
$z$-dependence of the amplitudes and frequency difference 
for $\Delta\beta=2.0$ (a) and $\Delta\beta=5.0$ (b). 
The initial conditions are $\eta_{1}(0)=1.001$, $\eta_{0}(0)=0.999$, 
and $\beta_{10}(0)=1.999$ in (a), and $\eta_{1}(0)=1.1$, 
$\eta_{0}(0)=0.9$, and $\beta_{10}(0)=5.1$ in (b). One observes that 
in (a) $\eta_{1}(z)$, $\eta_{0}(z)$, and $\beta_{10}(z)$ tend 
away from their equilibrium values $(1, 1, 2)$, while in (b) 
$\eta_{1}(z)$, $\eta_{0}(z)$, and $\beta_{10}(z)$  
oscillate about their equilibrium values $(1, 1, 5)$. 
Similar dynamic behavior is obtained for other initial conditions
in the neighborhood of $(1, 1, 2)$ or $(1, 1, 5)$.        
Based on these observations we conclude that the equilibrium state  
$(1, 1, 2)$ is unstable, whereas $(1, 1, 5)$ is stable, in agreement 
with linear stability analysis. This conclusion is 
further supported by the corresponding phase portraits that are 
shown in Fig. \ref{fig10}. For the system with $\Delta\beta=2.0$ 
the trajectories starting in the vicinity of $(1, 1, 2)$ tend away 
from $(1, 1, 2)$. In contrast, for the system with $\Delta\beta=5.0$ 
trajectories in the vicinity of $(1, 1, 5)$ are closed 
curves centered about $(1, 1, 5)$.

To further investigate the stability properties of a generic equilibrium 
point $(\eta, \eta, \Delta\beta)$ we perform detailed analysis of 
numerical simulations data for $\eta$ and $\Delta\beta$ values close to the 
bifurcation line predicted by linear stability calculations:   
$\Delta\beta_{bif}\simeq 2.31\eta^{1.5}$. Figure \ref{fig11} is 
the bifurcation diagram obtained by this analysis. 
It is seen that the bifurcation line obtained by numerical solution of 
Eq. (\ref{crosstalk37}) closely agrees with the line predicted by linear 
stability computations despite of the fact that $(\eta, \eta, \Delta\beta)$ 
is a non-hyperbolic equilibrium state. 
Notice that the points in the $\eta-\Delta\beta$ plane 
that are above the bifurcation line are stable, while those that are 
below the bifurcation line are unstable. Thus, 
transmission in the two-channel system described in subsection 
\ref{XPM} becomes unstable for frequency spacing values 
smaller than $2.31\eta^{1.5}$.

\subsection{Effects of Raman cross frequency shift}
\label{XFS}
The second effect of delayed Raman response on a single collision 
between a $j$th-channel soliton and a $k$th-channel soliton is 
an $O(\epsilon_{R}/|\beta_{j}-\beta_{k}|)$ frequency shift 
\cite{Chi89,Kumar98,Kaup99,P2004,CP2005,Agrawal96}. This 
Raman-induced XFS is given by \cite{CP2005,P2007,CP2008} 
\begin{eqnarray} &&
\Delta \beta_{j}=-\frac{8\epsilon_{R}\eta_{j}^{2}\eta_{k}}
{3|\beta_{k}-\beta_{j}|}.
\label{crosstalk40}
\end{eqnarray}     
We now consider the impact of this frequency shift on amplitude dynamics 
in a two-channel system. Since in general the amplitudes $\eta_{0}$ and 
$\eta_{1}$ vary with $z$, the frequency shifts might lead to $z$-dependence 
of the frequency difference $\beta_{10}$. In order to construct the perturbed 
model describing the effects of the Raman XFS we first obtain an 
equation for the dynamics of $\beta_{10}$. Using 
Eq. (\ref{crosstalk40}) we find 
$d\beta_{0}/dz=-16\epsilon_{R}\eta_{0}^{2}\eta_{1}/(3T)$ and     
$d\beta_{1}/dz=-16\epsilon_{R}\eta_{0}\eta_{1}^{2}/(3T)$. 
Combining these relations with the definition of 
$\beta_{10}$ we arrive at 
\begin{eqnarray} &&
\frac{d\beta_{10}}{dz}=
-16\epsilon_{R}\eta_{0}\eta_{1}\left(\eta_{1}-\eta_{0}\right)/(3T).
\label{crosstalk41}
\end{eqnarray}     
In addition, we replace the constant frequency difference 
$\Delta\beta$ by the $z$-dependent frequency difference 
$\beta_{10}(z)$ in the equations for $\eta_{0}$ and $\eta_{1}$ 
to obtain 
\begin{eqnarray} &&
\frac{d \eta_{1}}{dz}=
\eta_{1}\left(g_{1}-\frac{C}{\Delta\beta}\beta_{10}\eta_{0}\right),
\nonumber \\&&
\frac{d \eta_{0}}{dz}=
\eta_{0}\left(g_{0}+\frac{C}{\Delta\beta}\beta_{10}\eta_{1}\right).
\label{crosstalk42}
\end{eqnarray}      
Equations (\ref{crosstalk41}) and (\ref{crosstalk42}) represent 
the perturbed model for amplitude dynamics in the presence of the 
Raman XFS. 

We look for equilibrium states of the 
model in the form $(\eta, \eta, \Delta\beta)$ and find 
$g_{1}=-g_{0}=C\eta$ for the gain/loss coefficients.
Therefore, the Raman XFS does not modify the values of the gain/loss 
coefficients required for maintaining an equilibrium state with equal 
amplitudes. Substituting $g_{1}=-g_{0}=C\eta$ into Eq. (\ref{crosstalk42}) 
we obtain the following simpler form of the model:      
\begin{eqnarray} &&
\frac{d \eta_{1}}{dz}=
C\eta_{1}\left(\eta-\beta_{10}\eta_{0}/\Delta\beta\right),
\nonumber \\&&
\frac{d \eta_{0}}{dz}=
C\eta_{0}\left(-\eta+\beta_{10}\eta_{1}/\Delta\beta\right),
\nonumber \\&&
\frac{d\beta_{10}}{dz}=
-16\epsilon_{R}\eta_{0}\eta_{1}\left(\eta_{1}-\eta_{0}\right)/(3T).
\label{crosstalk43}
\end{eqnarray}     
It is possible to show that 
$V_{X}(\eta_{1}, \eta_{0}, \beta_{10})=8\eta_{0}\eta_{1}/3+\beta_{10}^{2}$ 
is a conserved quantity for the system (\ref{crosstalk43}). 
However, this does not guarantee the stability of the 
equilibrium point. Linear stability analysis predicts bifurcation 
at $\Delta\beta_{bif}=(8/3)^{1/2}\eta$. 
For $\Delta\beta>\Delta\beta_{bif}$ the Jacobian matrix of 
(\ref{crosstalk43}) has two purely imaginary eigenvalues and one zero 
eigenvalue, while for $\Delta\beta<\Delta\beta_{bif}$ 
all three eigenvalues are real. In the latter case one eigenvalue is positive, 
another is negative, and the third one is zero. 
For $\Delta\beta=\Delta\beta_{bif}$ all three eigenvalues 
are equal to zero. Based on this analysis one suspects that the 
equilibrium state $(\eta, \eta, \Delta\beta)$ becomes unstable for 
$\Delta\beta<\Delta\beta_{bif}$.

Since $(\eta, \eta, \Delta\beta)$  is a non-hyperbolic equilibrium 
point linear stability calculations might lead to erroneous conclusions 
\cite{Smale74,Wiggins91}. We therefore investigate the stability 
of $(\eta, \eta, \Delta\beta)$ by numerical simulations with Eq. 
(\ref{crosstalk43}). For concreteness we consider the two-channel 
system described in subsection \ref{XPM}.
Choosing $\eta=1$, the equilibrium state is $(1, 1, \Delta\beta)$ 
and the bifurcation value predicted by linear stability analysis 
is $\Delta\beta_{bif}\simeq 1.63$.    
Figure \ref{fig12} shows the dynamics of the amplitudes and frequency 
difference for $\Delta\beta=1.5$ (a) and $\Delta\beta=5.0$ (b). 
The initial conditions are $\eta_{1}(0)=1.001$, $\eta_{0}(0)=0.999$, 
and $\beta_{10}(0)=1.499$ in (a), and $\eta_{1}(0)=1.2$, 
$\eta_{0}(0)=0.9$, and $\beta_{10}(0)=5.05$ in (b). It is observed that 
in (a) $\eta_{1}(z)$, $\eta_{0}(z)$, and $\beta_{10}(z)$ tend 
away from their equilibrium values $(1, 1, 1.5)$, while in (b) 
$\eta_{1}(z)$, $\eta_{0}(z)$, and $\beta_{10}(z)$  
oscillate about the equilibrium values $(1, 1, 5)$. 
Additional numerical simulations with other initial conditions  
in the vicinity of $(1, 1, 1.5)$ or $(1, 1, 5)$ show the same 
dynamical behavior. We therefore conclude that the 
equilibrium state $(1, 1, 1.5)$ is unstable, while $(1, 1, 5)$ is stable,  
in agreement with linear stability analysis. This conclusion is also 
supported by the corresponding phase portraits.

To check the stability of a generic equilibrium state 
$(\eta, \eta, \Delta\beta)$ we carefully analyze results 
of numerical simulations with Eq. (\ref{crosstalk43}) for different 
$\eta$ and $\Delta\beta$ values. We pay special attention to the region 
in the $\eta$-$\Delta\beta$ plane that is in the close neighborhood 
of the bifurcation line $\Delta\beta_{bif}\simeq 1.63\eta$, predicted by 
linear stability computations. Figure \ref{fig13} shows the 
bifurcation diagram that is obtained by our analysis. 
It is seen that the bifurcation line obtained by numerical simulations 
is in good agreement with the line predicted by linear stability 
calculations, despite of the non-hyperbolic character of 
$(\eta, \eta, \Delta\beta)$. Notice that equilibrium points lying 
below the line $\Delta\beta = 1.63\eta$ are unstable, while 
those lying above it are stable. As a practical consequence we note that  
transmission in the two-channel system described in subsection 
\ref{XPM} is unstable for frequency spacing values 
smaller than $1.63\eta$.

\section{Conclusions}
\label{conclusions}
We studied the deterministic effects of inter-pulse Raman crosstalk in 
amplified WDM optical fiber transmission systems. We considered conventional 
optical solitons as an example for the pulses carrying the information   
and assumed that the pulse sequences in all frequency channels are 
deterministic and that the sequences are either infinitely long or 
are subject to periodic temporal boundary conditions. The first setup 
approximates return-to-zero (RZ) differential-phase-shift-keyed (DPSK)  
long-haul transmission, while the second one corresponds to RZ DPSK  
closed fiber loop experiments. We assumed in addition that the constant 
gain/loss in each frequency channel is determined by the difference between 
distributed amplifier gain and fiber loss. Under these assumptions we 
showed that the dynamics of pulse amplitudes in an 
$N$-channel transmission line is described by 
a system of $N$ coupled nonlinear ordinary differential equations (ODEs),  
having the form of an $N$-dimensional predator-prey model.   
We calculated the gain/loss coefficients required for maintaining an 
equilibrium state with equal non-zero amplitudes in all channels, and showed 
that high-frequency channels should be overamplified, while 
low-frequency channels should be underamplified compared with the 
middle (reference) channel. This means that the net gain/loss profile  
should {\it not} be flat with respect to the frequency. With these 
values of the gain/loss coefficients we proved stability of 
equilibrium states with non-zero amplitudes in all channels 
by constructing Lyapunov functions for the system of ODEs. 
The stability was found to be independent of the exact details 
of the approximation for the Raman gain curve. 
Furthermore, since the Lyapunov functions are conserved quantities for 
the system, typical dynamics of the amplitudes for initial conditions that 
are off the equilibrium points is oscillatory.

In an actual optical fiber line Raman crosstalk is not the only process 
impacting pulse dynamics. It is therefore important to understand the 
manner in which other physical processes perturb the Raman-induced amplitude 
dynamics described above. In this study we concentrated on the effects of 
three perturbations due to cross phase modulation (XPM), Raman self 
frequency shift (SFS) and Raman cross frequency shift (XFS). For each 
of these physical processes we constructed the corresponding perturbed model 
for an $N$-channel system and studied the dynamics in a two-channel system. 
For XPM-perturbed two-channel transmission we found that the gain/loss 
coefficients required for maintaining an equilibrium state with 
equal non-zero amplitudes are smaller compared with the values in the 
unperturbed case. This is explained by noting that the XPM-induced 
position shifts tend to increase the inter-collision distance and thus 
to decrease the rate of collisions. In contrast, the stability of the 
equilibrium state with equal non-zero amplitudes is not changed by XPM, i.e.,  
the equilibrium state remains a center. For two-channel systems perturbed 
by Raman SFS or Raman XFS we found that the values of the gain/loss 
coefficients required for maintaining the equilibrium state are the 
same as in the unperturbed case. However, the stability properties 
of the equilibrium state change as the frequency difference between 
the channels is decreased or increased, i.e., the system undergoes 
a bifurcation. The bifurcation curves are given by  
$\Delta\beta_{bif}=(16T\eta^{3}/15)^{1/2}$ for the perturbed 
model with Raman SFS and $\Delta\beta_{bif}=(8/3)^{1/2}\eta$ 
for the perturbed model with Raman XFS. In both models, for a fixed value 
of $\eta$, two-channel transmission with $\Delta\beta>\Delta\beta_{bif}$ is 
stable, while two-channel transmission with $\Delta\beta<\Delta\beta_{bif}$
is unstable. We therefore conclude that the Raman-induced interplay 
between amplitude dynamics and frequency dynamics sets a bound on 
the smallest frequency spacing for stable transmission.

In summary, our study provides a quantitative explanation to the stability 
of WDM DPSK transmission against Raman crosstalk effects. 
This stability was demonstrated in experiments 
in a closed fiber loop \cite{Golovchenko2000}. The stable behavior of 
Raman-induced amplitude dynamics in DPSK transmission is very different 
from the intermittent dynamic behavior exhibited by pulse parameters in 
on-off-keyed (OOK) transmission due to the interplay between Raman 
crosstalk and bit-pattern randomness \cite{Tkach95,Ho2000,P2007,CP2008,P2009}. 
This different dynamic behavior is an important advantage of 
DPSK transmission over OOK transmission.

\appendix
\section{Perturbed models in WDM transmission with $2N+1$ channels}
\label{appendB}
In this appendix we derive the perturbed models for Raman-induced 
amplitude dynamics in the presence of XPM and Raman SFS and XFS 
for WDM transmission lines with $2N+1$ channels. 

\subsection{Cross phase modulation}
The rate of collisions of a soliton from the $j$th 
channel with solitons from the the $k$th channel in the 
{\it unperturbed} model is 
\begin{eqnarray}&&
R_{jk}^{u}=|k-j|/\Delta z_{c}^{(1)}, 
\label{appendB1}
\end{eqnarray}
where $\Delta z_{c}^{(1)}=T/(2\Delta\beta)$ is constant. 
Thus, Eq. (\ref{crosstalk4}) for the change of the 
soliton amplitude in the interval $(z_{l-1},z_{l-1}+\Delta z_{c}^{(1)}]$
in the {\it unperturbed} model can be written as 
\begin{eqnarray} &&
\!\!\!\!\!\!\!\!\!\!\eta_{j}(z_{l-1}+\Delta z_{c}^{(1)})=
\eta_{j}(z_{l-1})+g_{j}\eta_{j}(z_{l-1})\Delta z_{c}^{(1)} 
\nonumber \\&&
+2\epsilon_{R}\Delta z_{c}^{(1)}\!\!\!\sum_{k=-N}^{N}
R_{jk}^{u}\mbox{sgn}(k-j)f(|j-k|)\eta_{j}(z_{l-1})\eta_{k}(z_{l-1}). 
\label{appendB2}
\end{eqnarray}       
In the {\it perturbed} transmission system, Eq. (\ref{appendB1}) is  
replaced by
\begin{eqnarray}&&
R_{jk}^{p}(z)=|k-j|/\Delta z_{cjk}^{(1)}(z), 
\label{appendB3}
\end{eqnarray}
where the $z$-dependent inter-collision distance $\Delta z_{cjk}^{(1)}$ 
is affected by the XPM-induced position shifts. Replacing $R_{jk}^{u}$ 
by $R_{jk}^{p}$ in Eq. (\ref{appendB2}) while employing relation 
(\ref{appendB3}) we arrive at       
\begin{eqnarray} &&
\!\!\!\!\!\!\!\!\!\!\eta_{j}(z_{l-1}+\Delta z_{c}^{(1)})=
\eta_{j}(z_{l-1})+g_{j}\eta_{j}(z_{l-1})\Delta z_{c}^{(1)} 
\nonumber \\&&
+2\epsilon_{R}\Delta z_{c}^{(1)}\!\!\!\sum_{k=-N}^{N}
\frac{{k-j}}{\Delta z_{cjk}^{(1)}(z_{l-1})}f(|j-k|)
\eta_{j}(z_{l-1})\eta_{k}(z_{l-1}). 
\label{appendB4}
\end{eqnarray}   
In order to find an expression for $\Delta z_{cjk}^{(1)}(z)$ we write 
down an equation for the location of the collision of the soliton from 
the zeroth time slot in the $j$th channel with the soliton from the $l(j-k)$ 
time slot in the $k$th channel: $y_{k,l(j-k)}(z_{l})=y_{j,0}(z_{l})$. 
Taking into account the different group velocities and summing over all 
XPM-induced position shifts during the collisions we arrive at the 
following generalization of Eq. (\ref{crosstalk28}): 
\begin{eqnarray}&&
\!\!\! y_{k,l(j-k)}(z_{l-1})+2k\Delta\beta\Delta z_{cjk}^{(1)}(z_{l-1})
+\frac{4}{(\Delta\beta)^{2}}\sum_{m=-N}^{N}\eta_{m}(z_{l-1})
\frac{|m-k|\mbox{sgn}(\beta_{m}-\beta_{k})(1-\delta_{mk})}
{(m-k)^{2}}
\nonumber\\&&
\!\!\!=y_{j,0}(z_{l-1})+2j\Delta\beta\Delta z_{cjk}^{(1)}(z_{l-1})
+\frac{4}{(\Delta\beta)^{2}}\sum_{m=-N}^{N}\eta_{m}(z_{l-1})
\frac{|m-j|\mbox{sgn}(\beta_{m}-\beta_{j})(1-\delta_{mj})}
{(m-j)^{2}},
\nonumber\\&&
\label{appendB5}
\end{eqnarray}
where $\delta_{ij}$ is the Kronecker delta function. Solution 
of Eq. (\ref{appendB5}) for  $\Delta z_{cjk}^{(1)}(z_{l-1})$ yields  
\begin{eqnarray}
\Delta z_{cjk}^{(1)}(z_{l-1})=\frac{T}{2\Delta\beta}
\left\{1+\frac{4}{T(j-k)(\Delta\beta)^{2}}
\sum_{m=-N}^{N}\eta_{m}(z_{l-1})
\left[\frac{1-\delta_{mk}}{m-k}-
\frac{1-\delta_{mj}}{m-j}\right]\right\}.
\label{appendB6}
\end{eqnarray}   
Taking the continuum limit in Eqs. (\ref{appendB4}) and 
(\ref{appendB6}) we obtain 
\begin{eqnarray} &&
\frac{d \eta_{j}}{dz}=
\eta_{j}\left[g_{j}+ 
2\epsilon_{R}\sum_{k=-N}^{N}\frac{{k-j}}{\Delta z_{cjk}^{(1)}}
f(|j-k|)\eta_{k}\right].  
\label{appendB7}
\end{eqnarray}        
Equation (\ref{appendB7}) together with Eq. (\ref{appendB6}) 
(with $z_{l-1}$ replaced by $z$) represent the perturbed model for 
Raman-induced amplitude dynamics in the presence of XPM in 
WDM transmission systems with $2N+1$ channels.

\subsection{Raman self and cross frequency shifts}
Consider the perturbed model with the Raman-induced SFS.  
The change in the amplitude of a $j$th-channel soliton  within the 
interval $(z_{l-1},z_{l-1}+\Delta z_{c}^{(1)}]$ is still given by 
Eq. (\ref{appendB4}). However, now the $z$-dependent inter-collision 
distance $\Delta z_{cjk}^{(1)}$ is given by   
\begin{eqnarray}&&
\Delta z_{cjk}^{(1)}(z_{l-1})=(k-j)T/(2\beta_{kj}(z_{l-1})), 
\label{appendB11}
\end{eqnarray}
where $\beta_{kj}(z)=\beta_{k}(z)-\beta_{j}(z)$.  
The dynamics of $\beta_{kj}$ is governed by  
\begin{eqnarray}&&
\frac{d\beta_{kj}}{dz}=-\frac{8}{15}\epsilon_{R}
\left(\eta_{k}^{4}-\eta_{j}^{4}\right).
\label{appendB12}
\end{eqnarray}     
Substituting relation (\ref{appendB11}) into Eq. (\ref{appendB4}) 
and going to the continuum limit we obtain 
\begin{eqnarray} &&
\frac{d \eta_{j}}{dz}=
\eta_{j}\left[g_{j}+
\frac{4\epsilon_{R}}{T}\sum_{k=-N}^{N}f(|j-k|)\beta_{kj}\eta_{k}\right].  
\label{appendB13}
\end{eqnarray}     
Equations (\ref{appendB12}) and (\ref{appendB13}) with $-N\le j,k \le N$ 
describe the Raman-induced amplitude dynamics in the presence of Raman SFS 
in transmission systems with $2N+1$ channels. It is straightforward to show 
that the gain/loss coefficients $g_{j}$ that are required for maintaining   
an equilibrium state with equal non-zero amplitudes in all channels are given 
by Eq. (\ref{crosstalk6}), that is, the values of these coefficients  
are not modified by the Raman SFS.

Turning to the perturbed model with Raman XFS we observe that 
amplitude dynamics is described by Eq. (\ref{appendB13}). To obtain 
the dynamic equation for the frequency difference $\beta_{kj}$ we 
first compute the change in the frequency experienced by a 
soliton in the $j$th channel within the interval 
$(z_{l-1},z_{l-1}+\Delta z_{c}^{(1)}]$. 
Employing Eq. (\ref{crosstalk40}) and summing over all collisions 
we arrive at 
\begin{eqnarray} &&
\!\!\!\!\!\!\!\!\!\!\beta_{j}(z_{l-1}+\Delta z_{c}^{(1)})=
\beta_{j}(z_{l-1})-\frac{8\epsilon_{R}}{3\Delta\beta}
\eta_{j}^{2}(z_{l-1})\sum_{k=-N}^{N}\eta_{k}(z_{l-1})(1-\delta_{kj}).
\label{appendB16}
\end{eqnarray}          
The continuum limit of Eq. (\ref{appendB16}) is 
\begin{eqnarray} &&
\frac{d \beta_{j}}{dz}=
-\frac{16\epsilon_{R}}{3T}\eta_{j}^{2}
\sum_{k=-N}^{N}\eta_{k}(1-\delta_{kj}),  
\label{appendB17}
\end{eqnarray}     
and therefore the dynamics of $\beta_{kj}(z)$ is governed by 
\begin{eqnarray} &&
\frac{d \beta_{kj}}{dz}=
-\frac{16\epsilon_{R}}{3T}
\left[\eta_{k}\eta_{j}(\eta_{k}-\eta_{j})+
\sum_{m=-N}^{N}\eta_{m}(\eta_{k}^{2}-\eta_{j}^{2})
(1-\delta_{mk})(1-\delta_{mj})\right].  
\label{appendB18}
\end{eqnarray}       
Thus, the perturbed model for Raman-induced amplitude dynamics 
in the presence of Raman XFS is given by Eqs. (\ref{appendB13}) 
and (\ref{appendB18}), where $-N\le j,k \le N$. 
A simple calculation shows that the values of the gain/loss coefficients 
required to maintain an equilibrium state with equal non-zero amplitudes 
in all frequency channels are the same as in the unperturbed model.

\newpage

\section*{List of Figure Captions}

\noindent Fig. 1. Amplitude dynamics for the two-channel system described 
by Eq. (\ref{simu1}). (a) The $z$-dependence of $\eta_{50}$ (solid curve) 
and $\eta_{0}$ (dotted curve) with the initial condition 
$\eta_{50}(0)=1.2$ and $\eta_{0}(0)=0.9$. 
(b) The corresponding phase portrait.

\noindent Fig. 2. Amplitude dynamics for the three-channel system described 
by Eq. (\ref{simu2}). (a) The phase portrait. (1,1,1) is the second 
equilibrium point from the right. 
(b) The $z$-dependence of $\eta_{50}$ (solid curve), 
$\eta_{0}$ (dotted curve), and $\eta_{-50}$ (dashed curve), with the 
initial condition  $\eta_{50}(0)=0.8$, $\eta_{0}(0)=0.9$, 
$\eta_{-50}(0)=1.1$.

\noindent Fig. 3. The $z$-dependence of pulse amplitudes for the four-channel 
system described by Eq. (\ref{simu7}) for $p=0$ (a), and $p=0.1$ (b). 
The initial condition is $\eta_{48}(0)=1.2$, $\eta_{16}(0)=1.1$, 
$\eta_{-16}(0)=0.95$, and $\eta_{-48}(0)=0.9$. The solid, dashed, 
dashed-dotted, and dotted curves represent $\eta_{48}(z)$, 
$\eta_{16}(z)$, $\eta_{-16}(z)$, and $\eta_{-48}(z)$, respectively. 

\begin{figure}[ptb]
\begin{tabular}{cc}
\epsfxsize=8.0cm  \epsffile{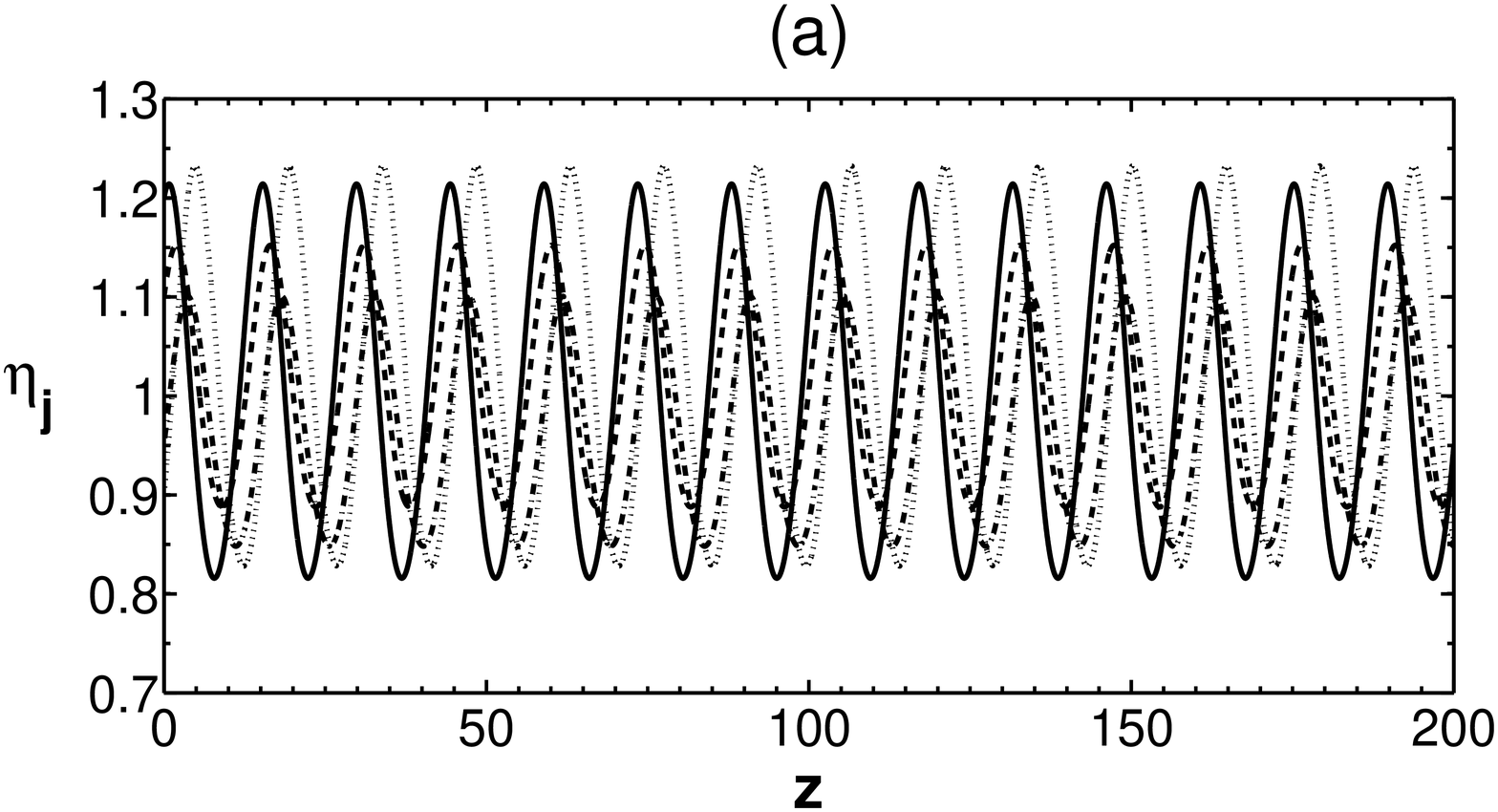} &
\epsfxsize=8.0cm  \epsffile{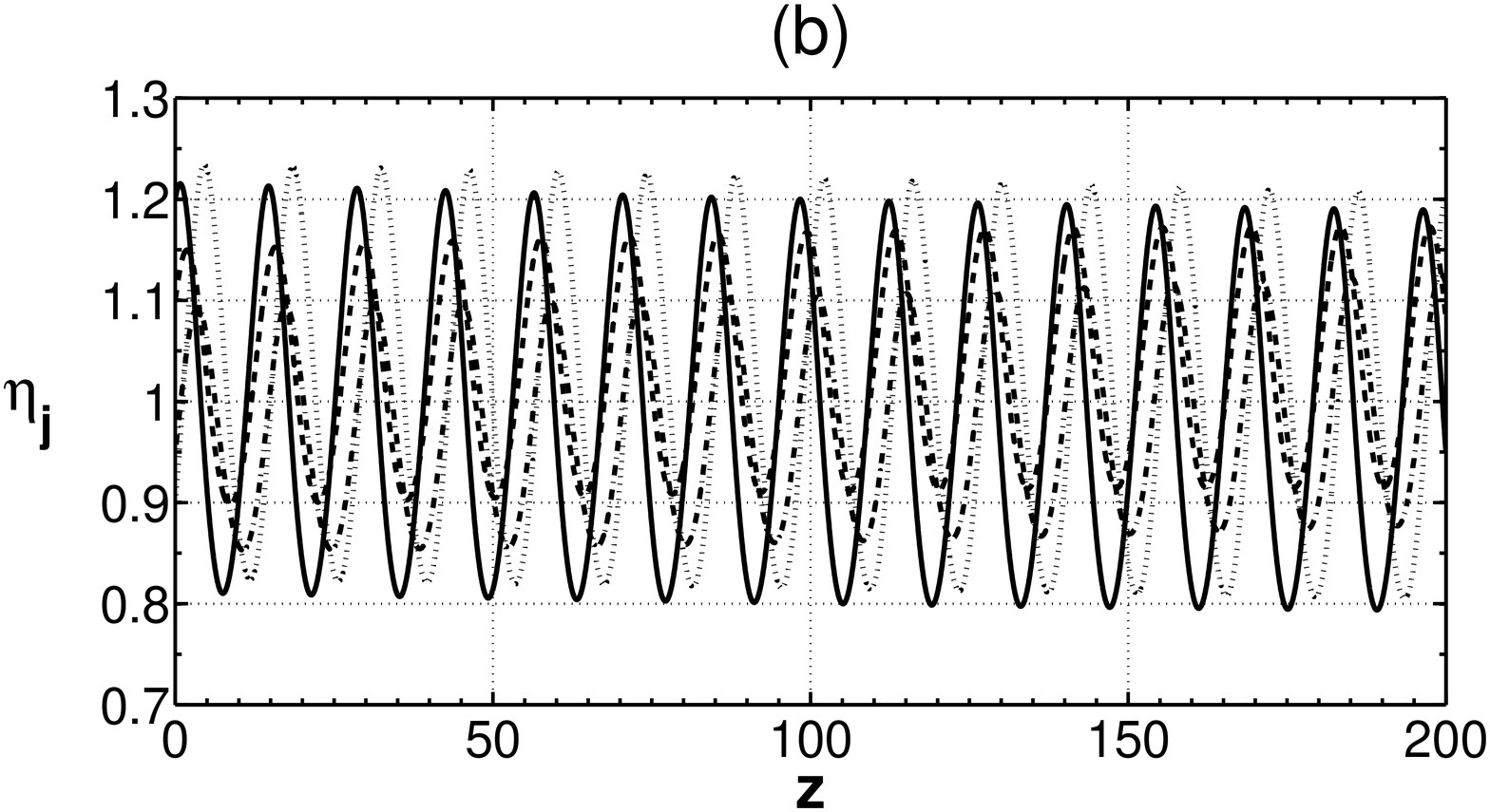} 
\end{tabular}
\caption{The $z$-dependence of pulse amplitudes for the four-channel 
system described by Eq. (\ref{simu7}) for $p=0$ (a), and $p=0.1$ (b). 
The initial condition is $\eta_{48}(0)=1.2$, $\eta_{16}(0)=1.1$, 
$\eta_{-16}(0)=0.95$, and $\eta_{-48}(0)=0.9$. The solid, dashed, 
dashed-dotted, and dotted curves represent $\eta_{48}(z)$, 
$\eta_{16}(z)$, $\eta_{-16}(z)$, and $\eta_{-48}(z)$, respectively.}
 \label{fig5}
\end{figure}

\begin{figure}[ptb]
\begin{tabular}{cc}
\epsfxsize=8.0cm  \epsffile{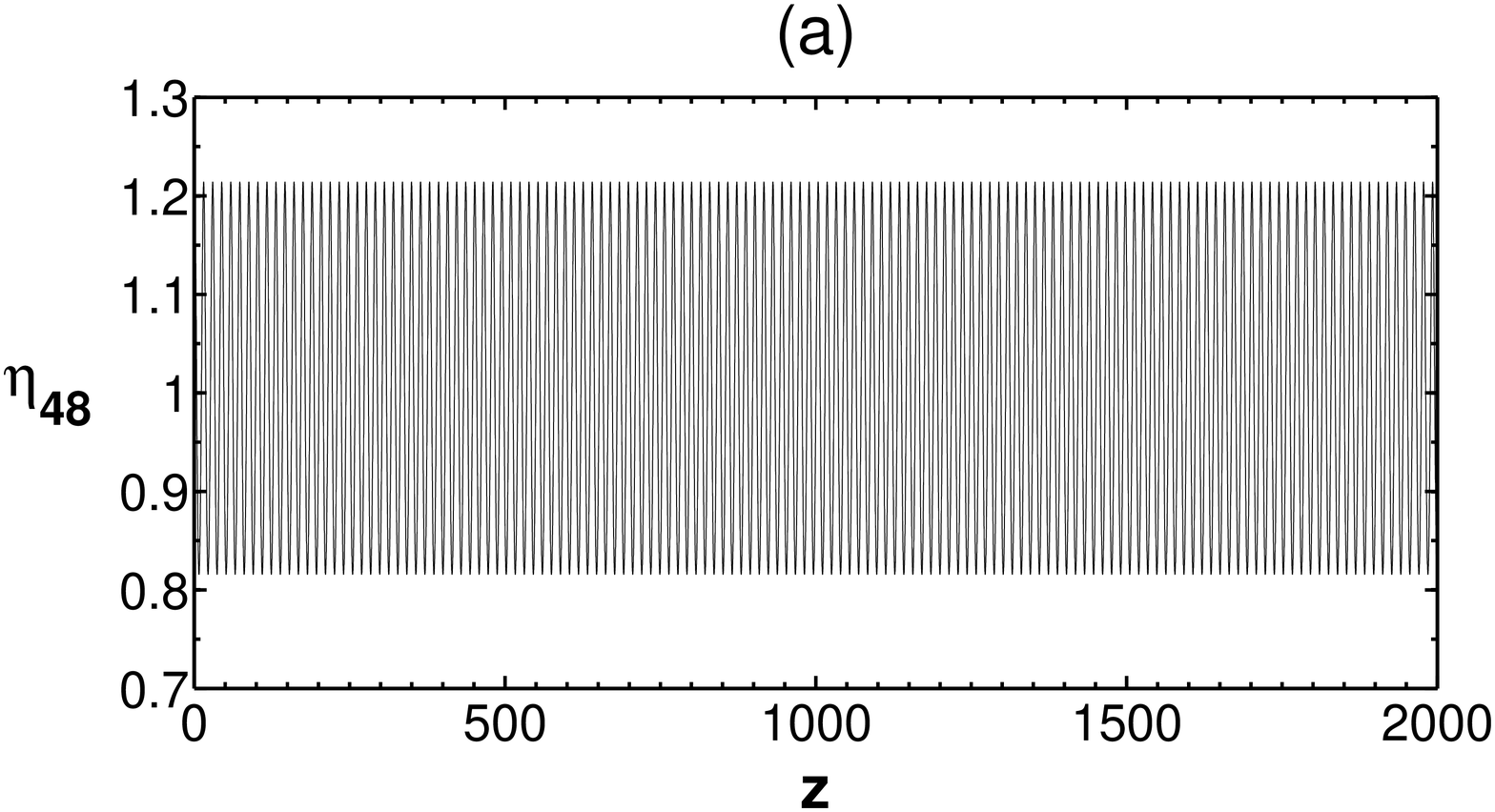} &
\epsfxsize=8.0cm  \epsffile{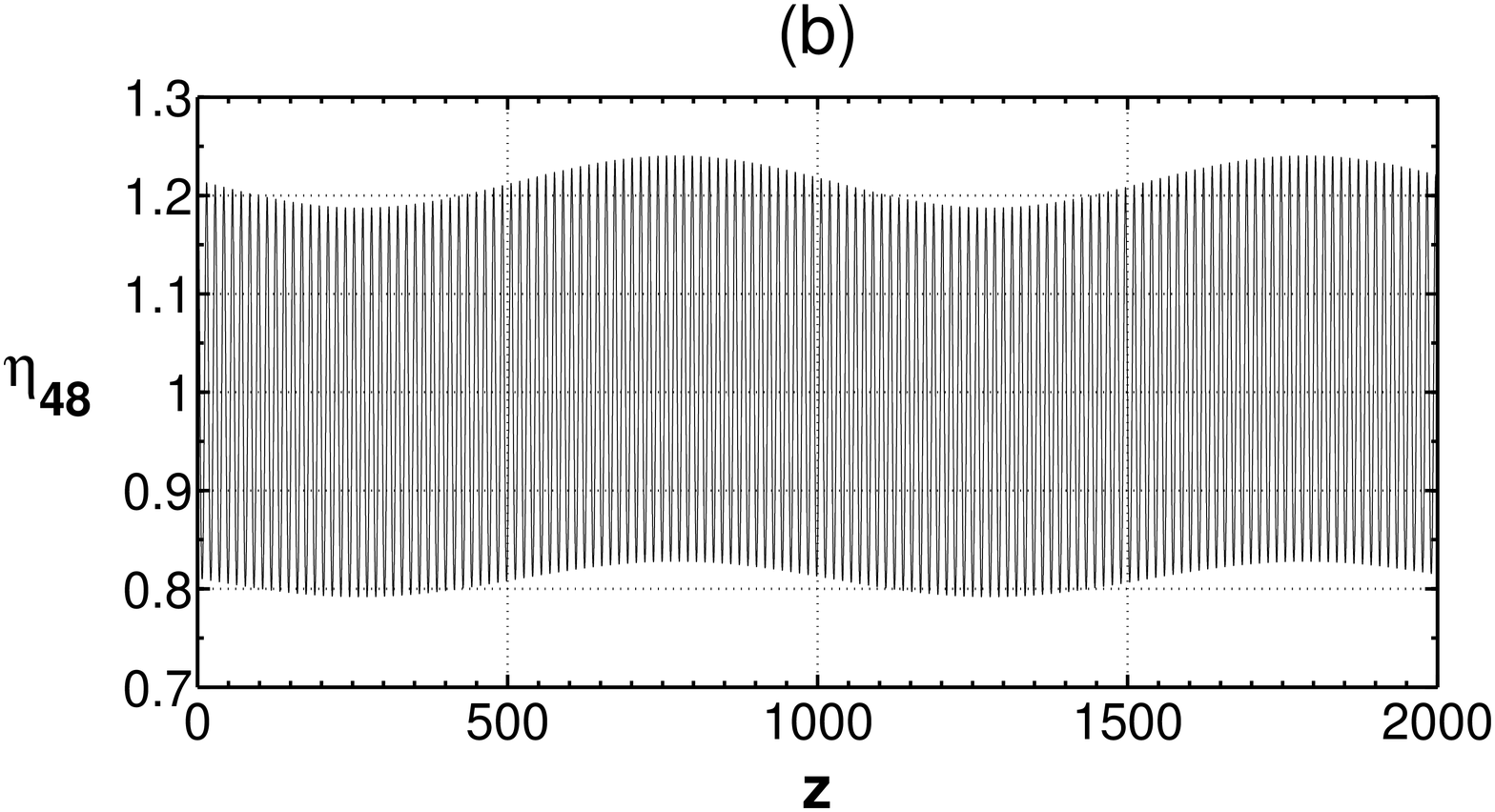} 
\end{tabular}
\caption{The $z$-dependence of the amplitudes of pulses in the $j=48$ 
channel for $p=0$ (a), and $p=0.1$ (b). The initial condition is the same as 
in Fig \ref{fig5} and the final propagation distance is $z_{f}=2000$.}
 \label{fig6}
\end{figure}

\begin{figure}[ptb]
\begin{tabular}{cc}
\epsfxsize=8.0cm  \epsffile{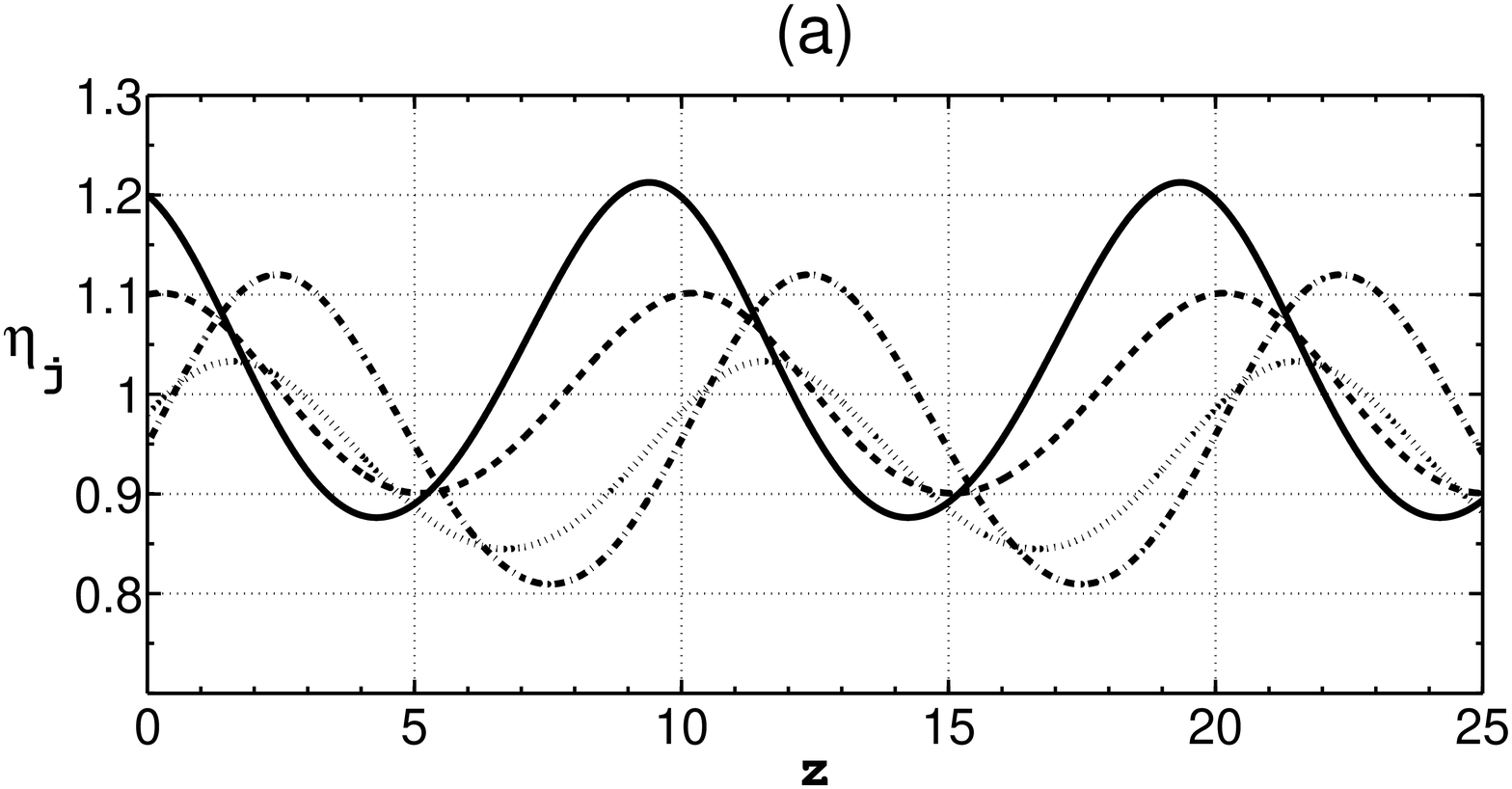} &
\epsfxsize=8.0cm  \epsffile{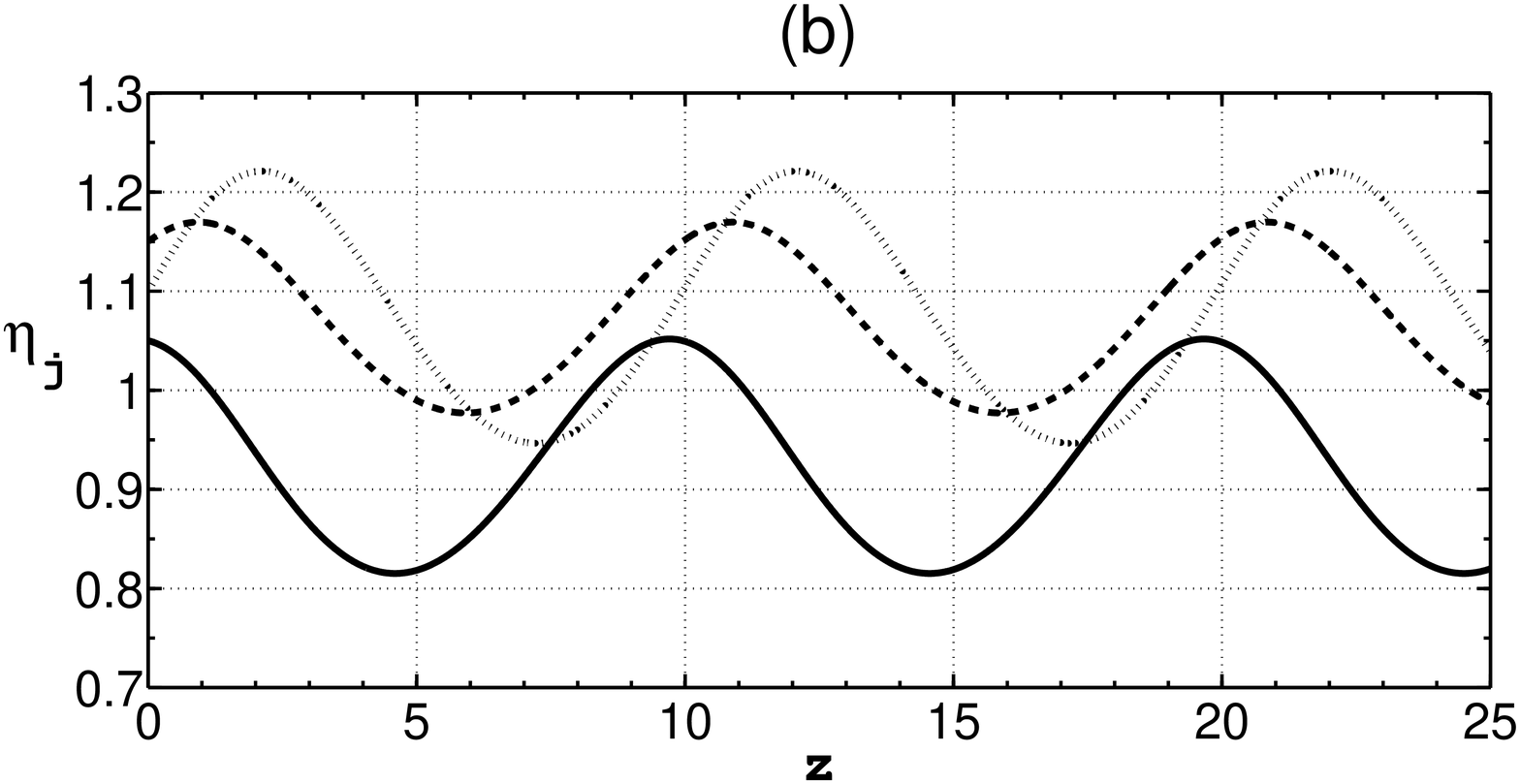} 
\end{tabular}
\caption{The $z$-dependence of pulse amplitudes for a seven-channel 
system operating at 10 Gbits/s per channel. The initial amplitudes are 
$\eta_{180}(0)=1.2$, $\eta_{120}(0)=1.05$, $\eta_{60}(0)=1.1$ 
$\eta_{0}(0)=1.15$, $\eta_{-60}(0)=0.98$, $\eta_{-120}(0)=1.1$, 
and $\eta_{-180}(0)=0.95$. The solid, dashed, dotted, 
and dashed-dotted curves in (a) represent $\eta_{180}(z)$, 
$\eta_{60}(z)$, $\eta_{-60}(z)$, and $\eta_{-180}(z)$, respectively. 
The solid, dashed, and dotted curves in (b) correspond to 
$\eta_{120}(z)$, $\eta_{0}(z)$, and $\eta_{-120}(z)$.}
 \label{7channels}
\end{figure}

\begin{figure}[ptb]
\begin{tabular}{cc}
\epsfxsize=8.0cm  \epsffile{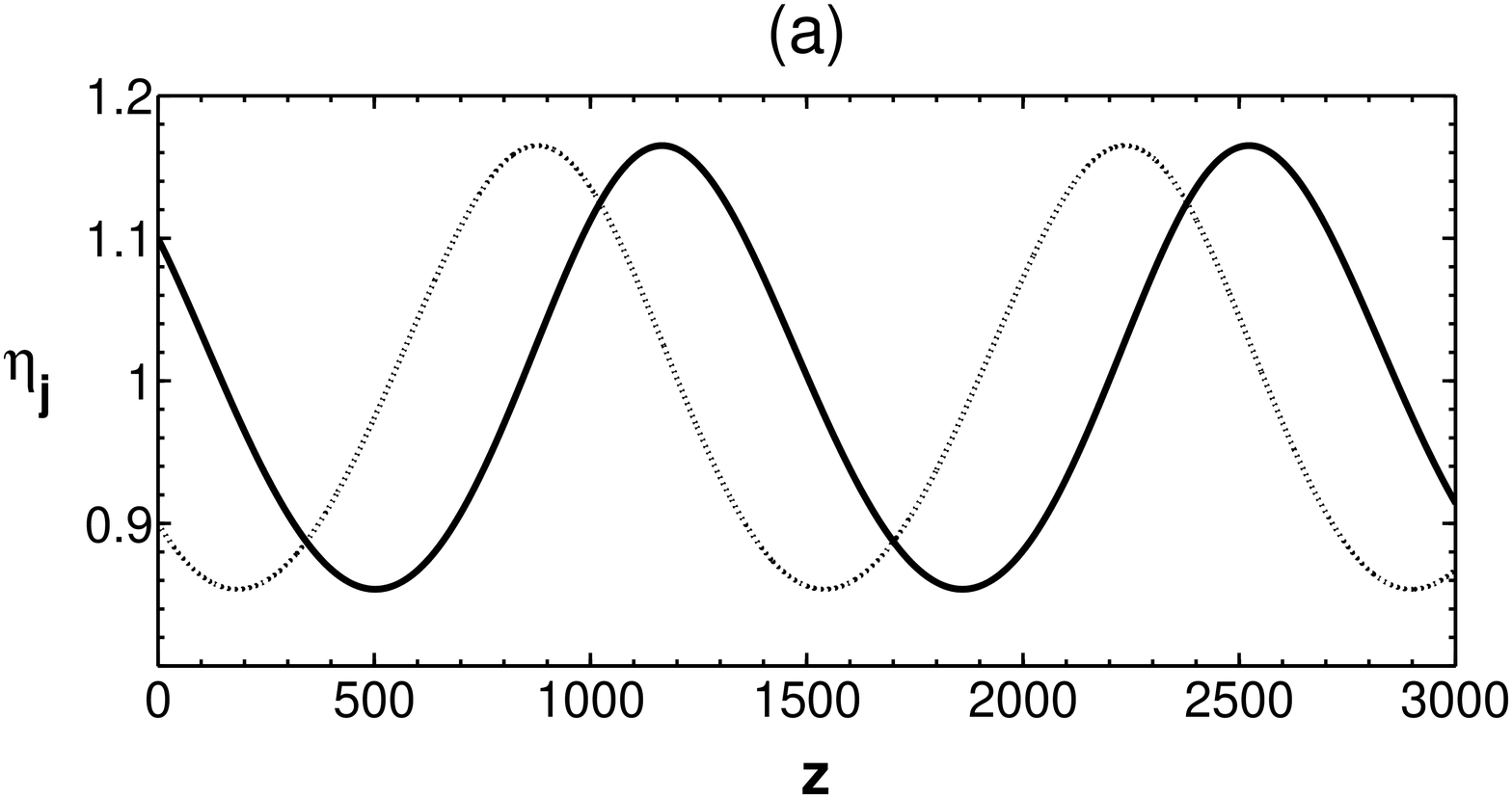} &
\epsfxsize=8.0cm  \epsffile{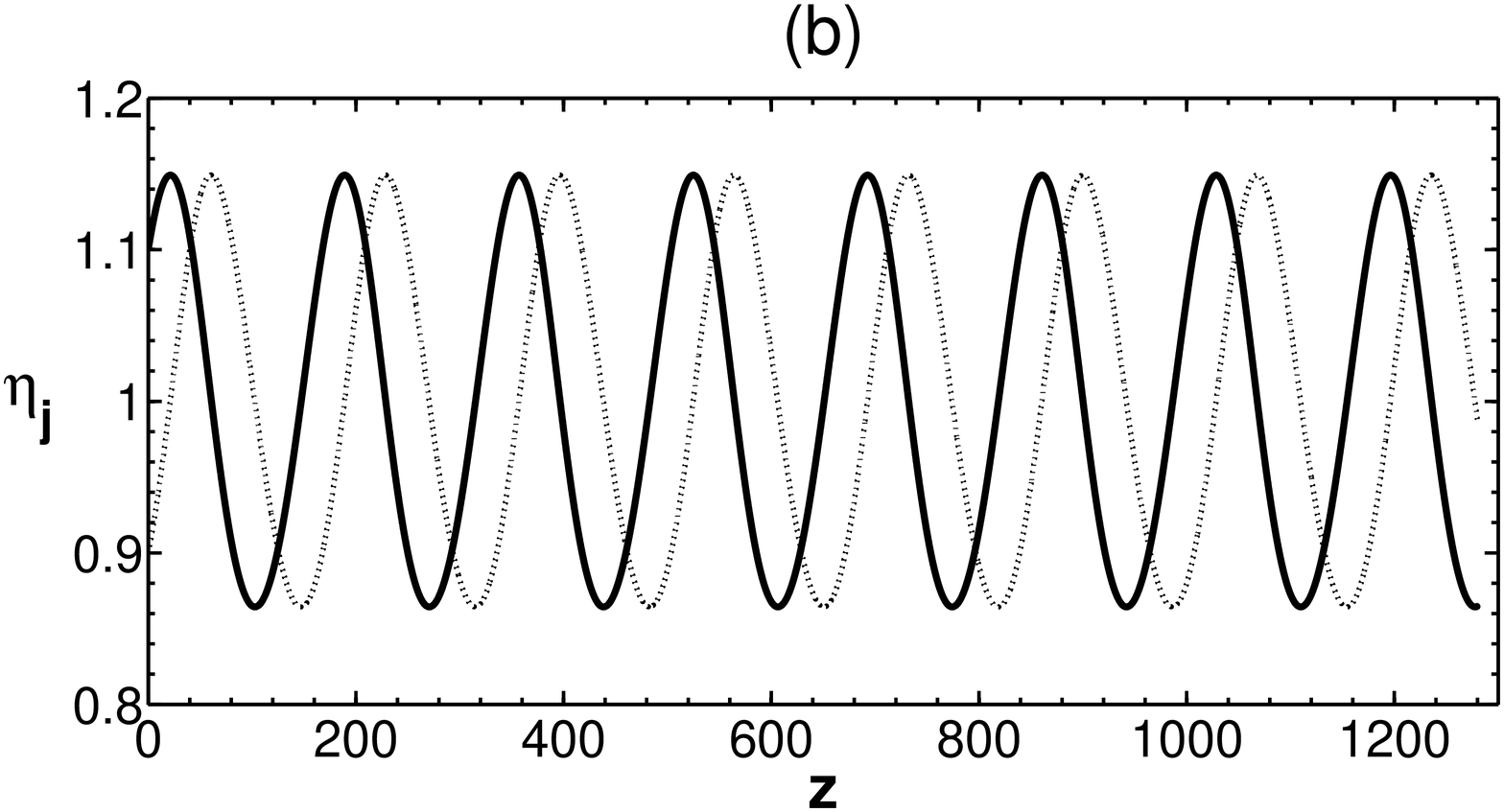} 
\end{tabular}
\caption{The $z$-dependence of pulse amplitudes for the XPM-perturbed 
two-channel system described by Eq. (\ref{crosstalk32}) with 
$\Delta\beta=2.0$ (a) and $\Delta\beta=10.0$ (b).  
The initial condition is $\eta_{1}(0)=1.1$ and $\eta_{0}(0)=0.9$. 
The solid and dotted lines represent $\eta_{1}(z)$, 
and $\eta_{0}(z)$, respectively.}
 \label{fig7}
\end{figure}

\begin{figure}[ptb]
\begin{tabular}{cc}
\epsfxsize=8.0cm  \epsffile{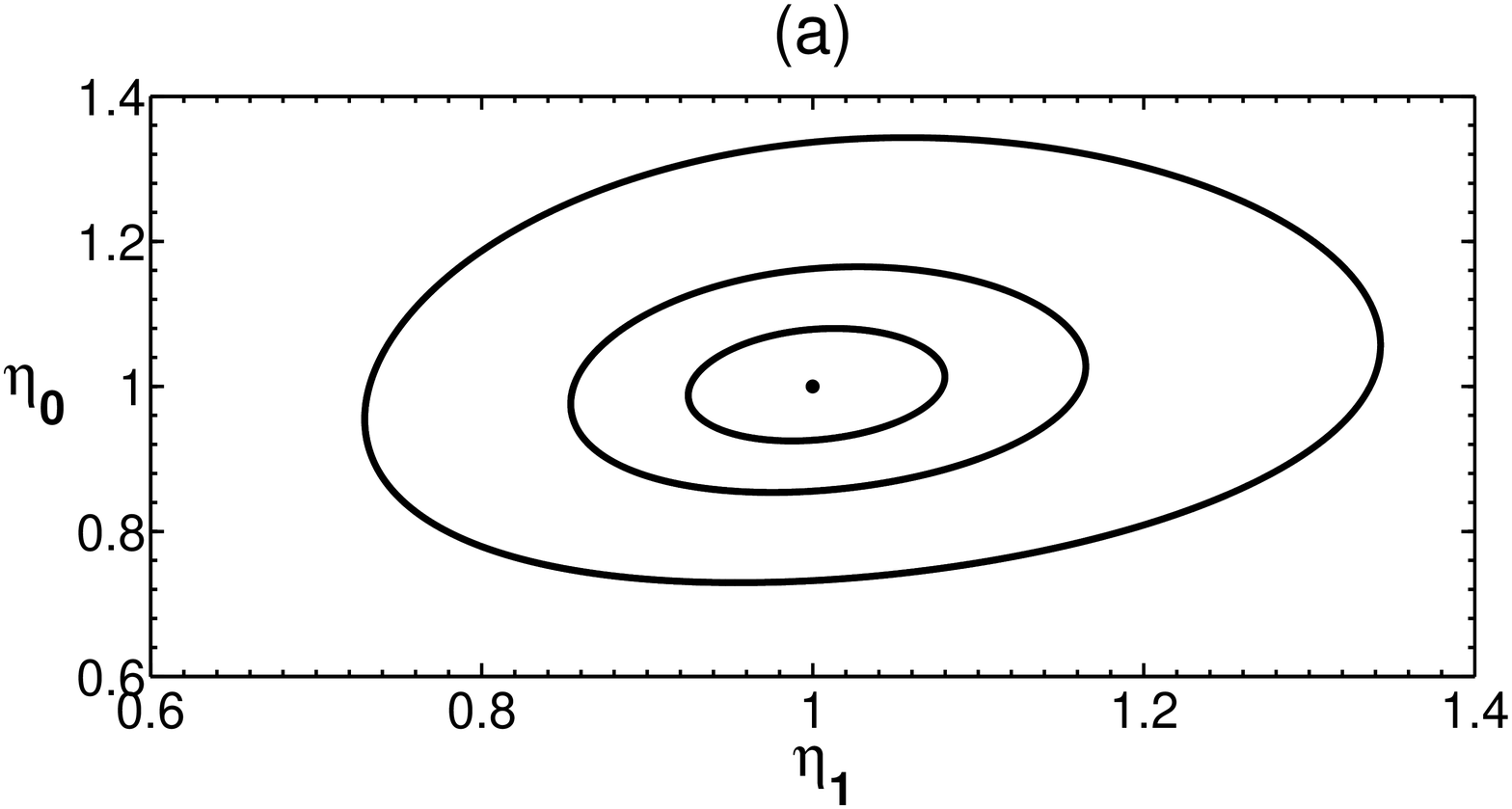} &
\epsfxsize=8.0cm  \epsffile{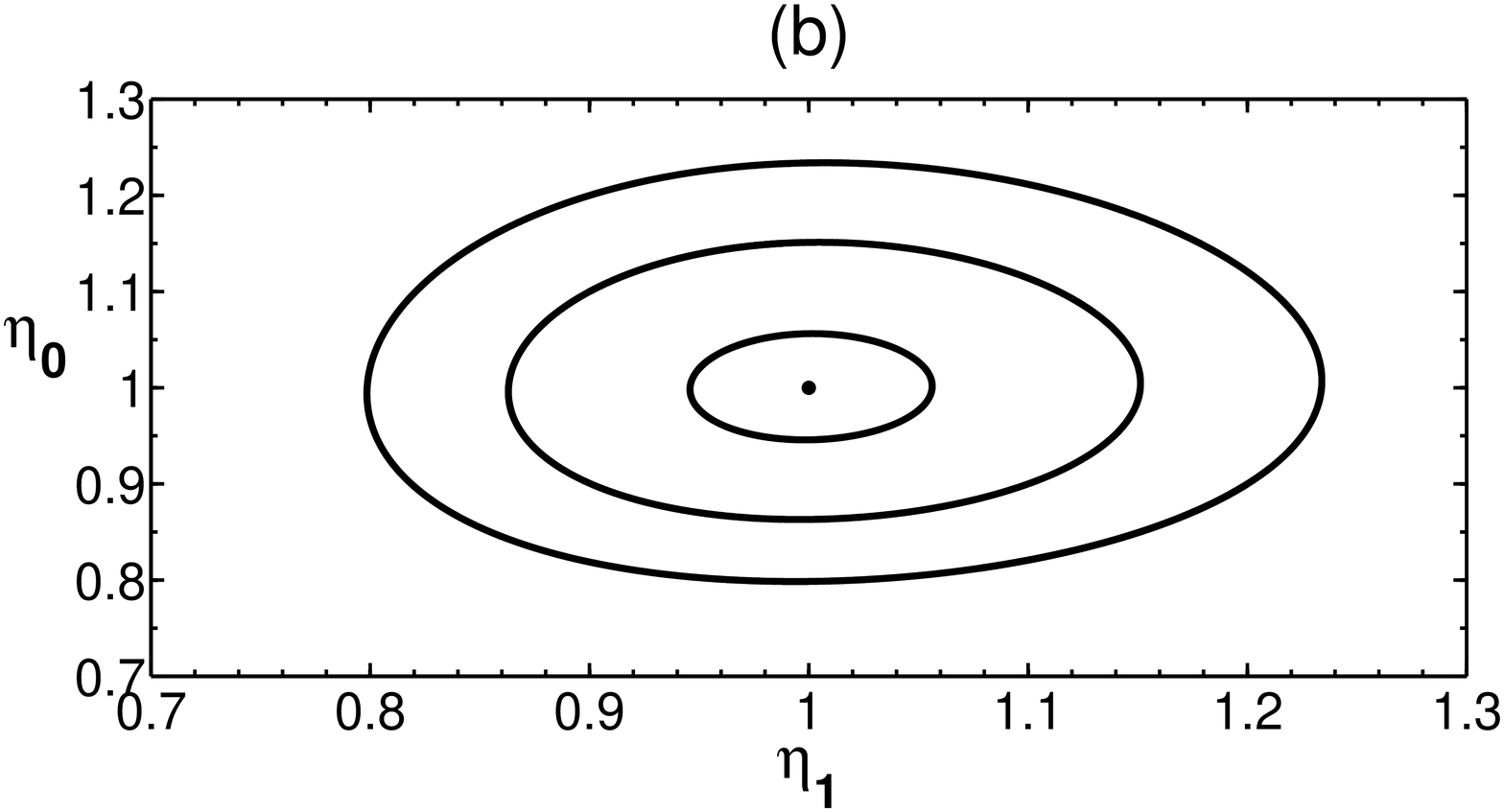} 
\end{tabular}
\caption{The phase portrait for the XPM-perturbed two-channel  
system described by Eq. (\ref{crosstalk32}) with 
$\Delta\beta=2.0$ (a) and $\Delta\beta=10.0$ (b).}
 \label{fig8}
\end{figure}

\begin{figure}[ptb]
\begin{tabular}{cc}
\epsfxsize=8.0cm  \epsffile{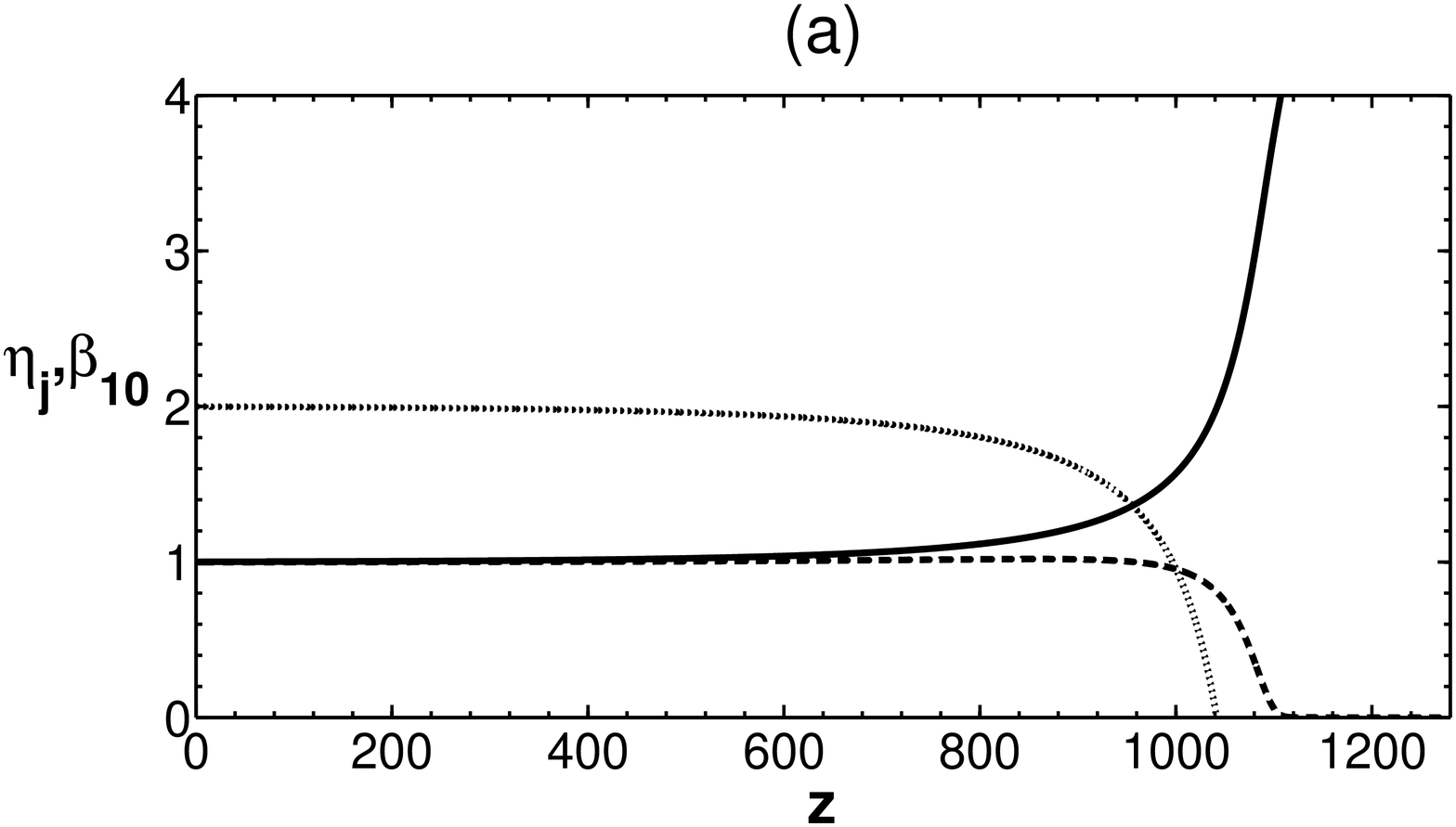} &
\epsfxsize=8.0cm  \epsffile{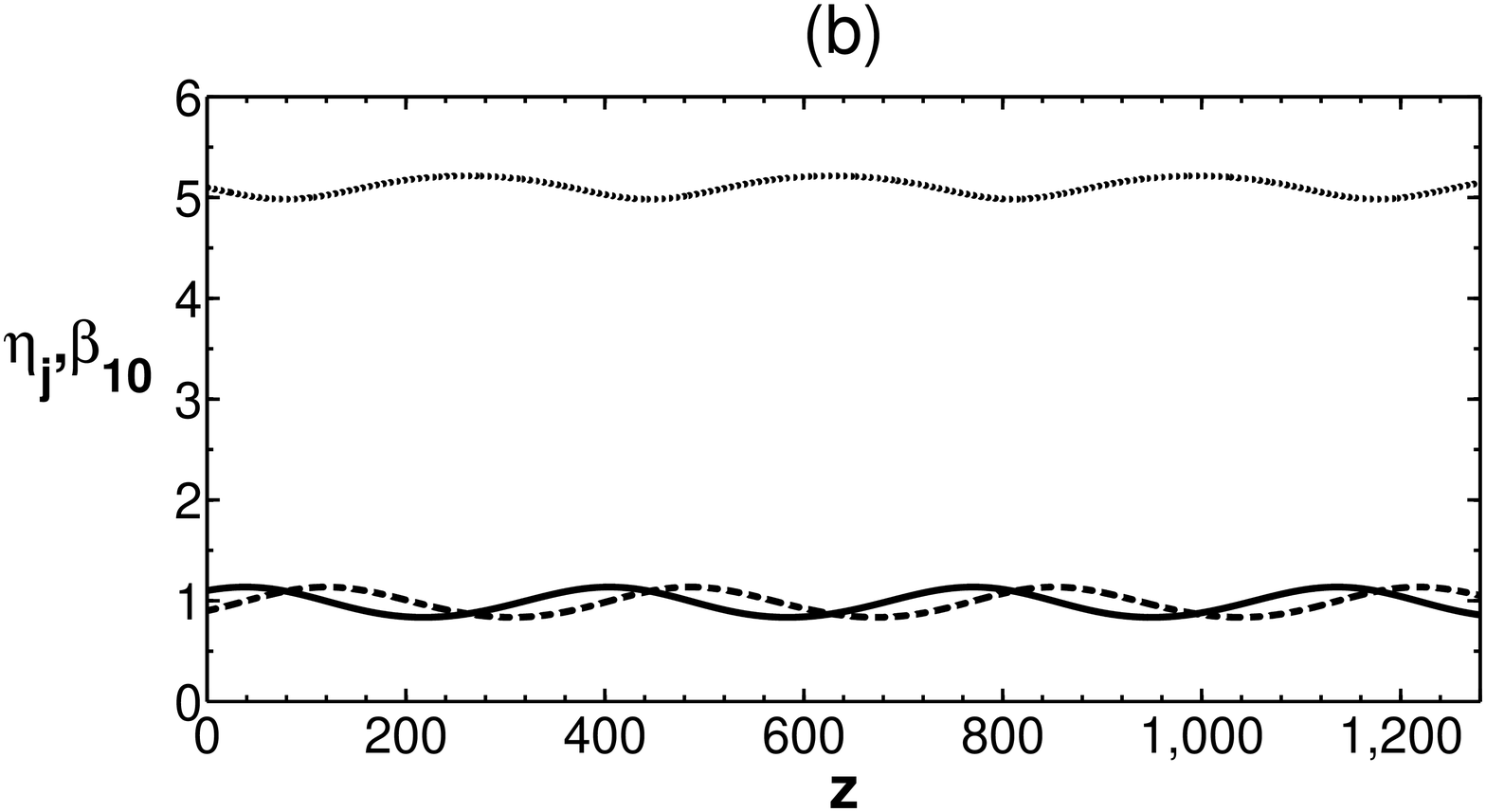} 
\end{tabular}
\caption{The $z$-dependence of soliton amplitudes and frequency 
difference for a two-channel system perturbed by the Raman 
SFS [Eq. (\ref{crosstalk37})].   
(a) The dynamics with $\Delta\beta=2.0$ and initial condition 
$\eta_{1}(0)=1.001$, $\eta_{0}(0)=0.999$, and $\beta_{10}(0)=1.999$.  
(b) The dynamics with $\Delta\beta=5.0$ and initial condition 
$\eta_{1}(0)=1.1$, $\eta_{0}(0)=0.9$, $\beta_{10}(0)=5.1$. 
The solid, dashed and dotted lines represent $\eta_{1}(z)$, 
$\eta_{0}(z)$, and $\beta_{10}(z)$, respectively.}
 \label{fig9}
\end{figure}

\begin{figure}[ptb]
\begin{tabular}{cc}
\epsfxsize=8.0cm  \epsffile{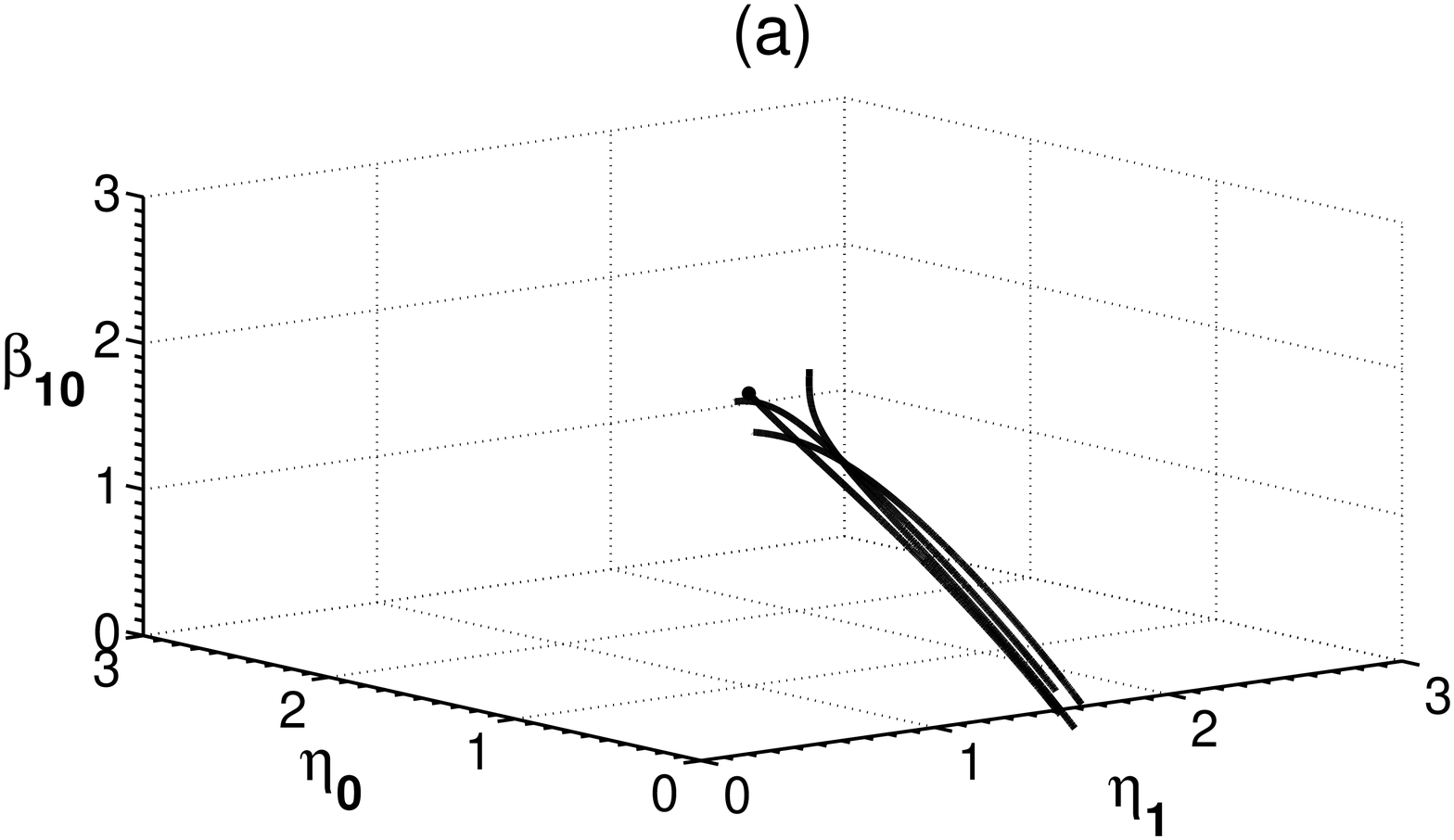} &
\epsfxsize=8.0cm  \epsffile{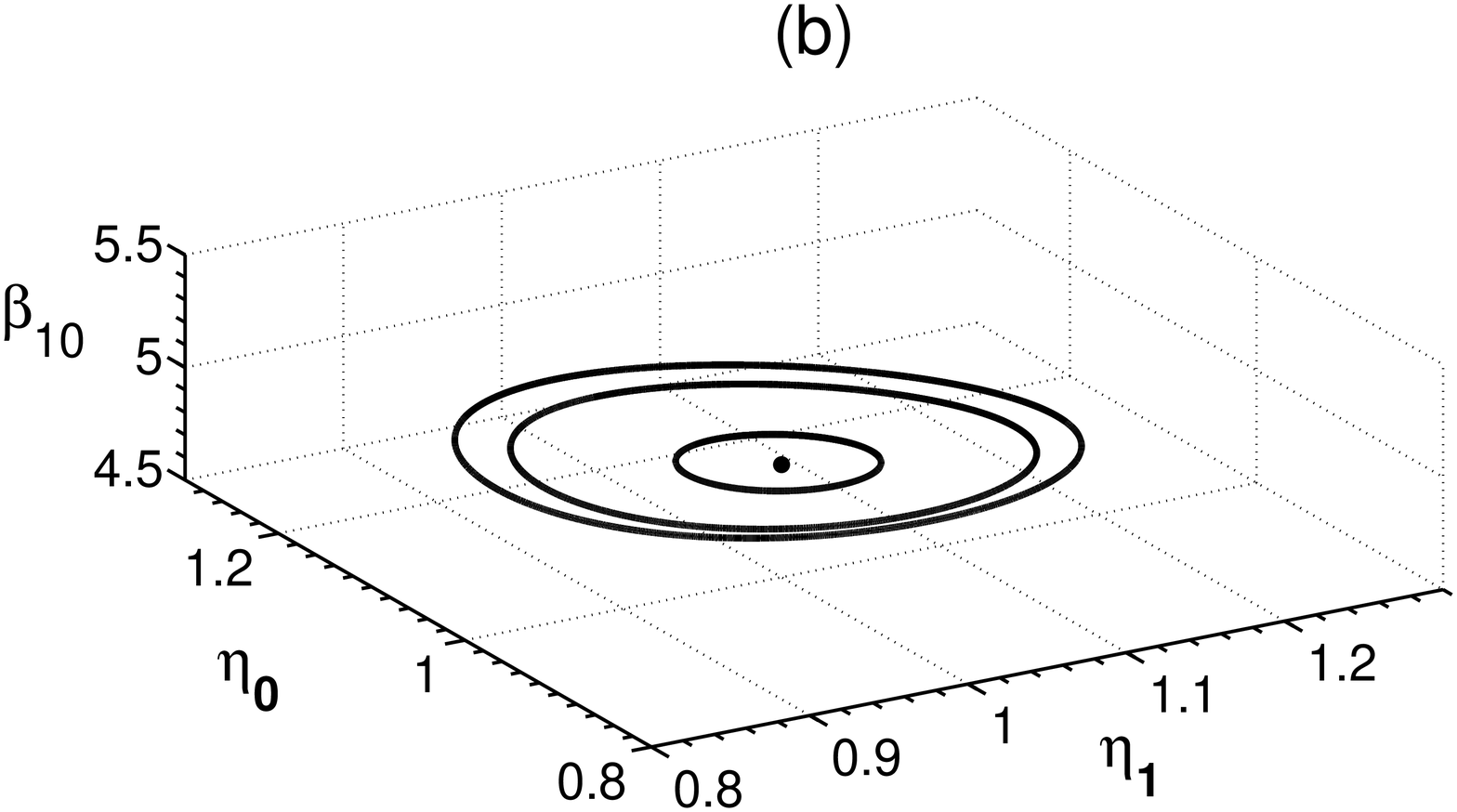} 
\end{tabular}
\caption{The phase portraits for the two-channel system perturbed by the 
Raman SFS with $\Delta\beta=2.0$ (a) and $\Delta\beta=5.0$ (b).}
 \label{fig10}
\end{figure}

\begin{figure}[ptb]
\begin{tabular}{cc}
\epsfxsize=12.0cm  \epsffile{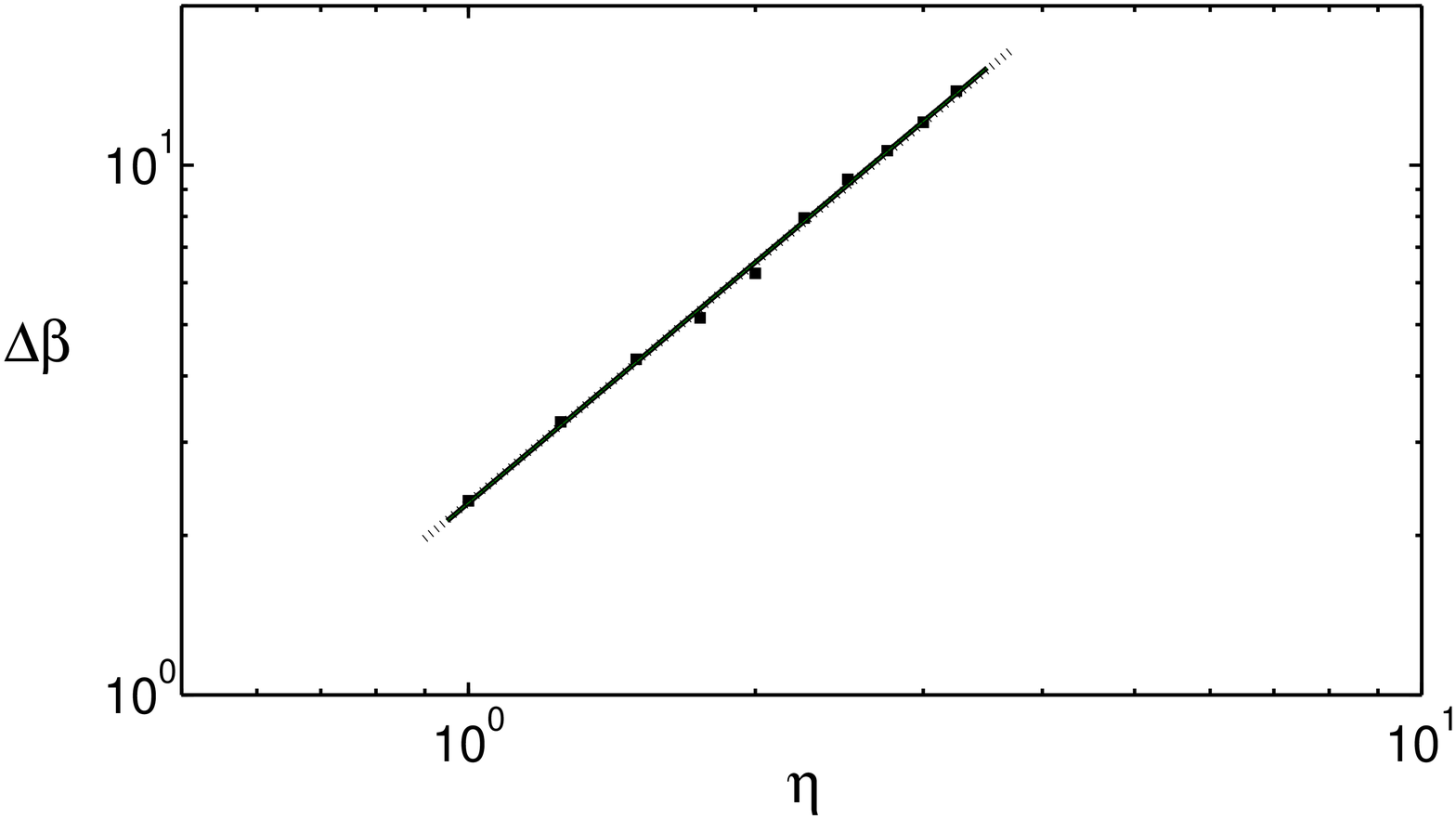} 
\end{tabular}
\caption{The bifurcation diagram for the two-channel system perturbed 
by the Raman SFS. The squares correspond to the bifurcation values obtained 
by numerical solution of Eq. (\ref{crosstalk37}), while the solid line 
is a fit of the form $\Delta\beta_{bif}=2.30\eta^{1.51}$ for the 
numerical data. The dotted line represents the prediction of linear 
stability analysis: $\Delta\beta_{bif}\simeq 2.31\eta^{1.5}$.}
 \label{fig11}
\end{figure}

\begin{figure}[ptb]
\begin{tabular}{cc}
\epsfxsize=8.0cm  \epsffile{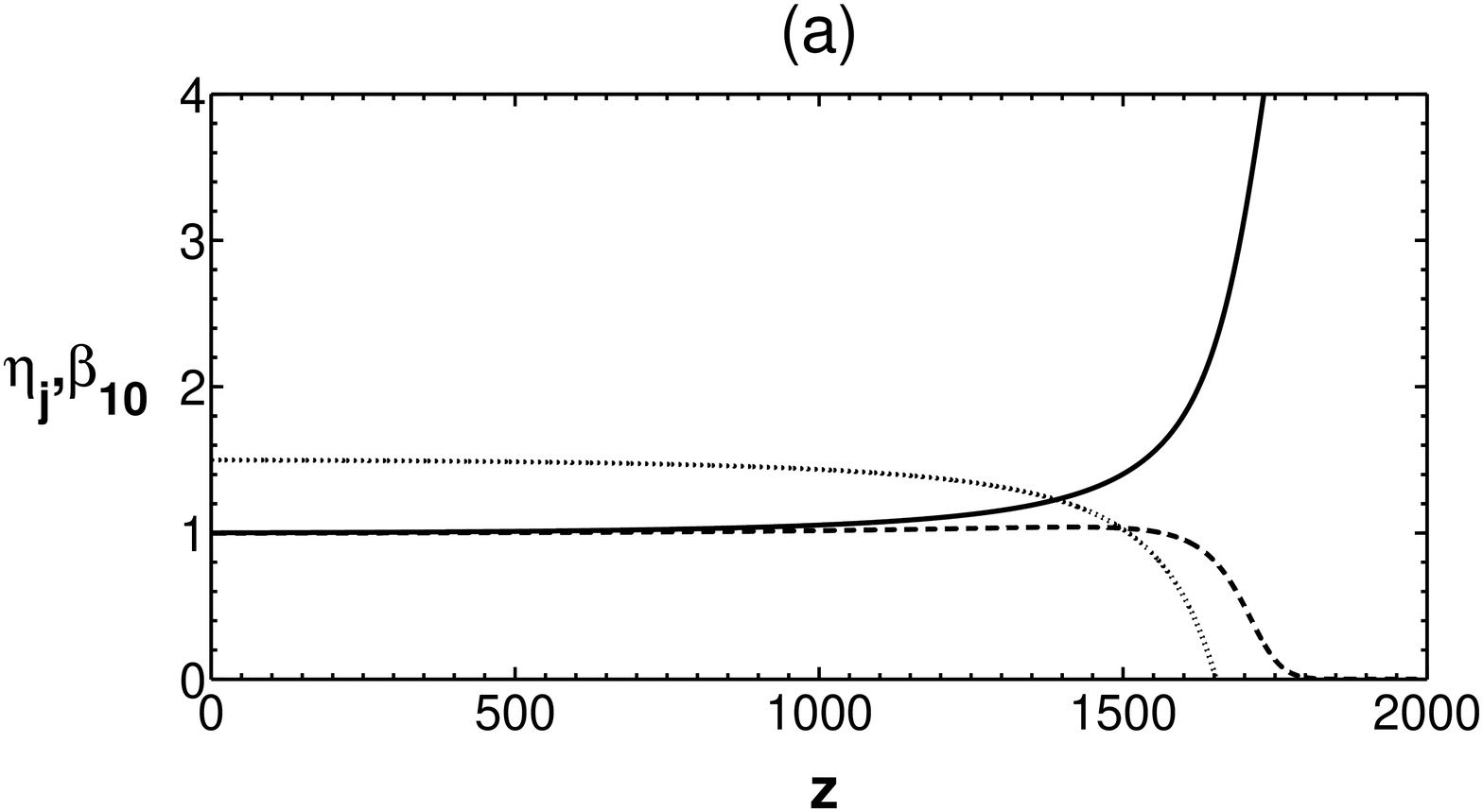} &
\epsfxsize=8.0cm  \epsffile{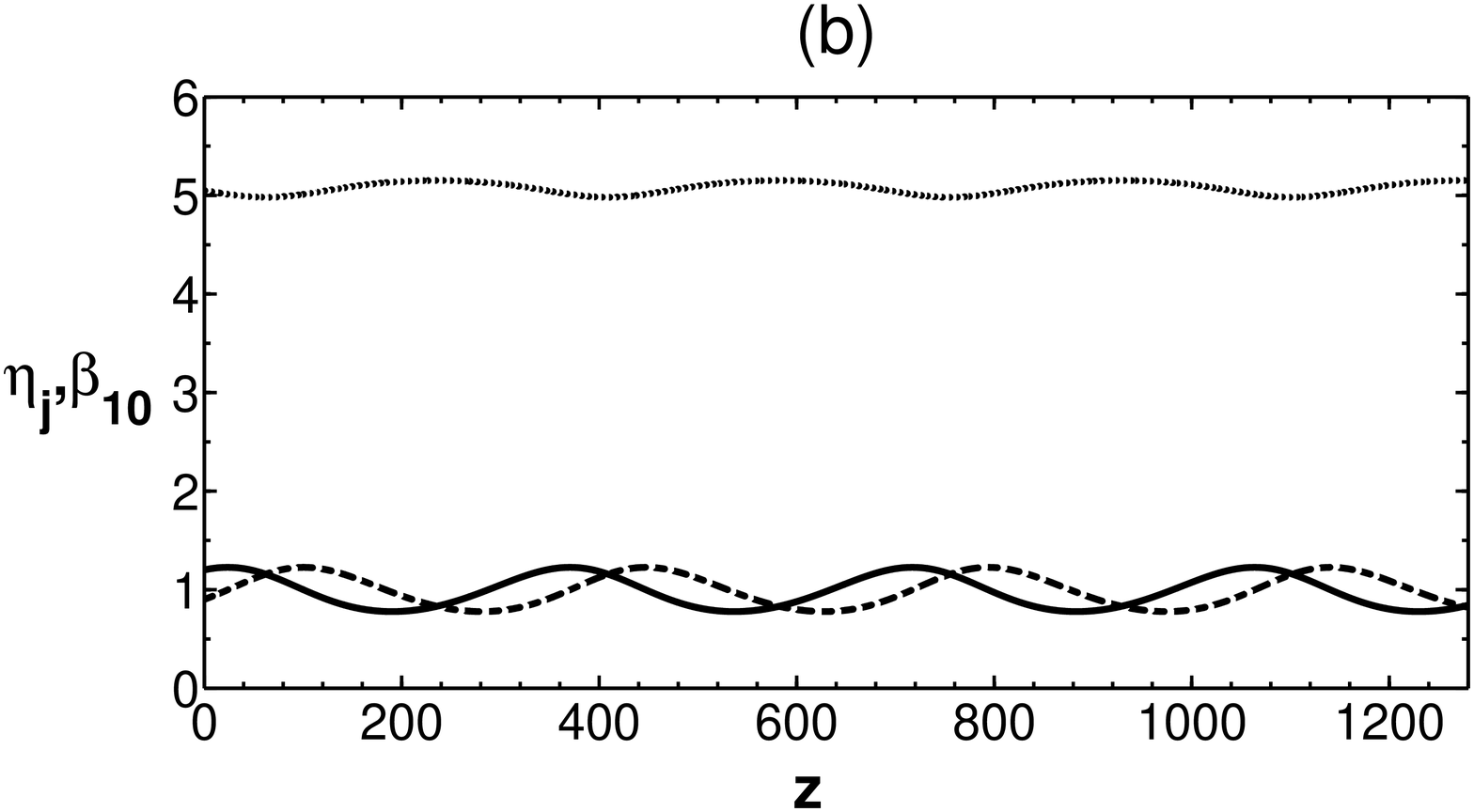} 
\end{tabular}
\caption{The $z$-dependence of soliton amplitudes and frequency 
difference for a two-channel system perturbed by 
the Raman XFS [Eq. (\ref{crosstalk43})].   
(a) The dynamics with $\Delta\beta=1.5$ and initial condition 
$\eta_{1}(0)=1.001$, $\eta_{0}(0)=0.999$, and $\beta_{10}(0)=1.499$.  
(b) The dynamics with $\Delta\beta=5.0$ and initial condition 
$\eta_{1}(0)=1.2$, $\eta_{0}(0)=0.9$, $\beta_{10}(0)=5.05$. 
The solid, dashed and dotted lines represent $\eta_{1}(z)$, 
$\eta_{0}(z)$, and $\beta_{10}(z)$, respectively.}
 \label{fig12}
\end{figure}

\begin{figure}[ptb]
\begin{tabular}{cc}
\epsfxsize=12.0cm  \epsffile{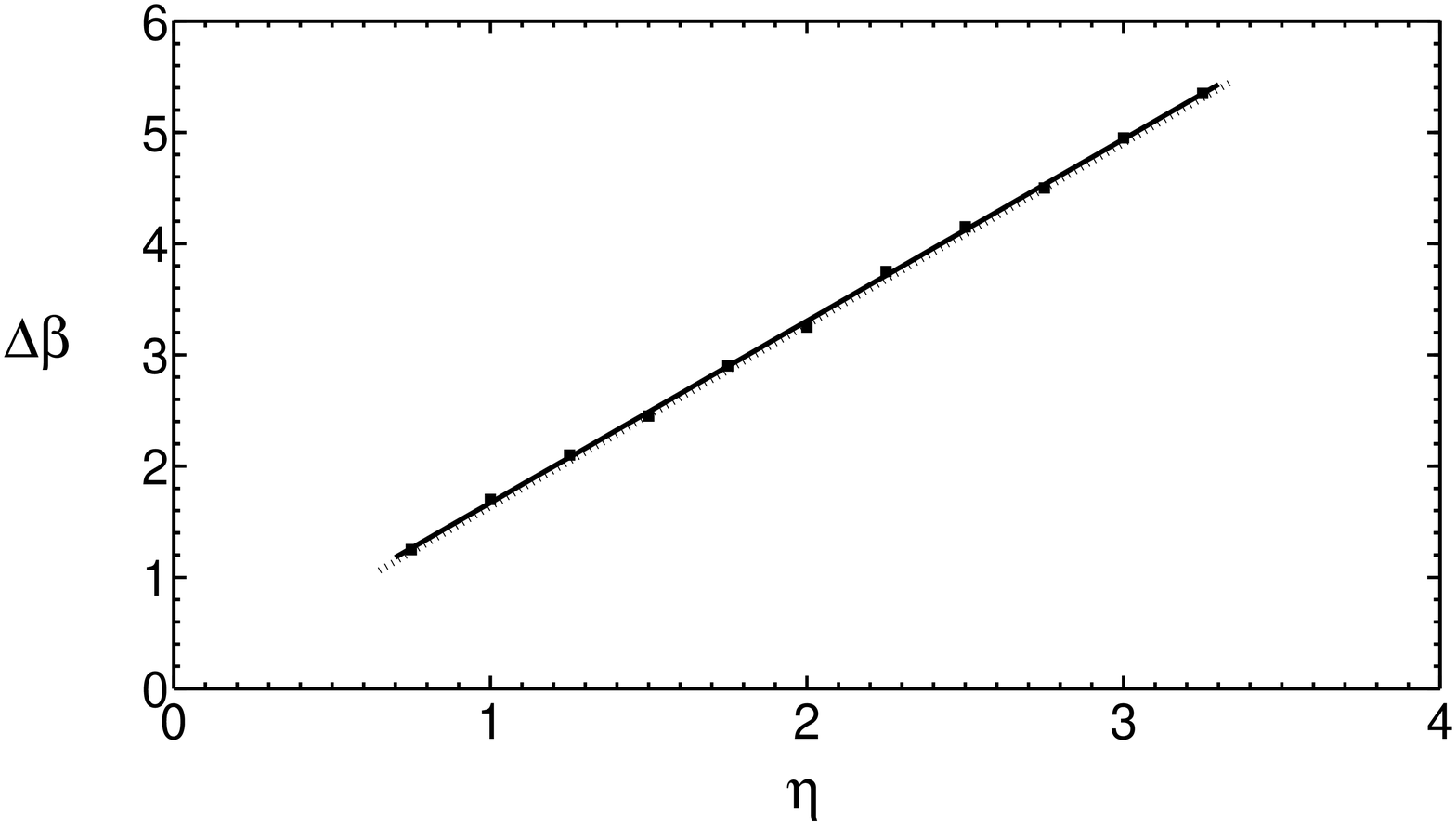} 
\end{tabular}
\caption{The bifurcation diagram for the two-channel system perturbed 
by the Raman XFS. The squares correspond to the bifurcation values obtained 
by numerical solution of Eq. (\ref{crosstalk43}), while the solid line 
is a fit of the form $\Delta\beta_{bif}=0.04+1.64\eta$ for the 
numerical data. The dotted line is the prediction of linear 
stability analysis: $\Delta\beta_{bif}\simeq 1.63\eta$.}
 \label{fig13}
\end{figure}

\noindent Fig. 4. The $z$-dependence of the amplitudes of pulses in the $j=48$ 
channel for $p=0$ (a), and $p=0.1$ (b). The initial condition is the same as 
in Fig \ref{fig5} and the final propagation distance is $z_{f}=2000$.

\noindent Fig. 5. The $z$-dependence of pulse amplitudes for a seven-channel 
system operating at 10 Gbits/s per channel. The initial amplitudes are 
$\eta_{180}(0)=1.2$, $\eta_{120}(0)=1.05$, $\eta_{60}(0)=1.1$ 
$\eta_{0}(0)=1.15$, $\eta_{-60}(0)=0.98$, $\eta_{-120}(0)=1.1$, 
and $\eta_{-180}(0)=0.95$. The solid, dashed, dotted, 
and dashed-dotted curves in (a) represent $\eta_{180}(z)$, 
$\eta_{60}(z)$, $\eta_{-60}(z)$, and $\eta_{-180}(z)$, respectively. 
The solid, dashed, and dotted curves in (b) correspond to 
$\eta_{120}(z)$, $\eta_{0}(z)$, and $\eta_{-120}(z)$.

\noindent Fig. 6. The $z$-dependence of pulse amplitudes for the XPM-perturbed 
two-channel system described by Eq. (\ref{crosstalk32}) with 
$\Delta\beta=2.0$ (a) and $\Delta\beta=10.0$ (b).  
The initial condition is $\eta_{1}(0)=1.1$ and $\eta_{0}(0)=0.9$. 
The solid and dotted lines represent $\eta_{1}(z)$, 
and $\eta_{0}(z)$, respectively. 

\noindent Fig. 7. The phase portrait for the XPM-perturbed two-channel  
system described by Eq. (\ref{crosstalk32}) with $\Delta\beta=2.0$ (a) 
and $\Delta\beta=10.0$ (b). 

\noindent Fig. 8. The $z$-dependence of soliton amplitudes and frequency 
difference for a two-channel system perturbed by the Raman 
SFS [Eq. (\ref{crosstalk37})].   
(a) The dynamics with $\Delta\beta=2.0$ and initial condition 
$\eta_{1}(0)=1.001$, $\eta_{0}(0)=0.999$, and $\beta_{10}(0)=1.999$.  
(b) The dynamics with $\Delta\beta=5.0$ and initial condition 
$\eta_{1}(0)=1.1$, $\eta_{0}(0)=0.9$, $\beta_{10}(0)=5.1$. 
The solid, dashed and dotted lines represent $\eta_{1}(z)$, 
$\eta_{0}(z)$, and $\beta_{10}(z)$, respectively.

\noindent Fig. 9. The phase portraits for the two-channel system 
perturbed by the Raman SFS with $\Delta\beta=2.0$ (a) and 
$\Delta\beta=5.0$ (b). 

\noindent Fig. 10. The bifurcation diagram for the 
two-channel system perturbed by the Raman SFS. 
The squares correspond to the bifurcation values obtained 
by numerical solution of Eq. (\ref{crosstalk37}), while the solid line 
is a fit of the form $\Delta\beta_{bif}=2.30\eta^{1.51}$ for the 
numerical data. The dotted line represents the prediction of linear 
stability analysis: $\Delta\beta_{bif}\simeq 2.31\eta^{1.5}$.

\noindent Fig. 11. The $z$-dependence of soliton amplitudes and frequency 
difference for a two-channel system perturbed by 
the Raman XFS [Eq. (\ref{crosstalk43})].   
(a) The dynamics with $\Delta\beta=1.5$ and initial condition 
$\eta_{1}(0)=1.001$, $\eta_{0}(0)=0.999$, and $\beta_{10}(0)=1.499$.  
(b) The dynamics with $\Delta\beta=5.0$ and initial condition 
$\eta_{1}(0)=1.2$, $\eta_{0}(0)=0.9$, $\beta_{10}(0)=5.05$. 
The solid, dashed and dotted lines represent $\eta_{1}(z)$, 
$\eta_{0}(z)$, and $\beta_{10}(z)$, respectively.

\noindent Fig. 12. The bifurcation diagram for the two-channel 
system perturbed by the Raman XFS. The squares correspond to the 
bifurcation values obtained by numerical solution of 
Eq. (\ref{crosstalk43}), while the solid line 
is a fit of the form $\Delta\beta_{bif}=0.04+1.64\eta$ for the 
numerical data. The dotted line is the prediction of linear 
stability analysis: $\Delta\beta_{bif}\simeq 1.63\eta$.

\newpage

\newpage

\newpage

\newpage

\newpage

\newpage

\newpage

\newpage

\newpage

\newpage

\newpage

\newpage

\begin{thebibliography}{}
\bibitem{Agrawal2001} G. P. Agrawal, {\it Nonlinear 
Fiber Optics} (Academic, San Diego, CA, 2001).

\bibitem{Stolen80} R. H. Stolen, Proc. IEEE {\bf 68}, 1232 (1980).     

\bibitem{Chraplyvy84}  A. R. Chraplyvy, Electron. Lett. {\bf 20}, 58 (1984).

\bibitem{Tkach95} F. Forghieri, R. W. Tkach, and A. R. Chraplyvy, 
IEEE Photon. Technol. Lett. {\bf 7}, 101 (1995).

\bibitem{Jander96} D. N. Christodoulides and R. B. Jander, 
IEEE Photon. Technol. Lett. {\bf 8}, 1722 (1996).   

\bibitem{Ho2000} K.-P. Ho, J. Lightwave Technol. {\bf 18}, 915 (2000).

\bibitem{Chi89} S. Chi and S. Wen, Opt. Lett. {\bf 14}, 1216 (1989).

\bibitem{Malomed91a} B. A. Malomed, Phys. Rev. A {\bf 44}, 1412 (1991).

\bibitem{Kumar98} S. Kumar, Opt. Lett. {\bf 23}, 1450 (1998).  

\bibitem{P2004} A. Peleg, Opt. Lett. {\bf 29}, 1980 (2004).

\bibitem{Kaup99} T. I. Lakoba and D. J. Kaup, Opt. Lett. {\bf 24}, 
808 (1999).

\bibitem{Cotter84} D. Cotter and A. M. Hill, 
Electron. Lett. {\bf 20}, 185 (1984).  

\bibitem{Sarkar2007} A. R. Sarkar, M. N. Islam and M.G. Mostafa, 
Opt. Quantum Electron. {\bf 39}, 659 (2007).

\bibitem{Kumar2003} M. Muktoyuk and S. Kumar, 
IEEE Photon. Technol. Lett. {\bf 15}, 1222 (2003).

\bibitem{CP2005} Y. Chung and A. Peleg, Nonlinearity {\bf 18}, 
1555 (2005). 

\bibitem{Yamamoto2003} T. Yamamoto and S. Norimatsu, 
J. Lightwave Technol. {\bf 21}, 2229 (2003).


\bibitem{P2007} A. Peleg, Phys. Lett. A {\bf 360}, 533 (2007).

\bibitem{CP2008} Y. Chung and A. Peleg, Phys. Rev. A  {\bf 77}, 063835 (2008). 

\bibitem{P2009} A. Peleg, Phys. Lett. A {\bf 373}, 2734 (2009). 

\bibitem{Zhou2006} X. Zhou and M. Birk, 
J. Lightwave Technol. {\bf 24}, 1218 (2006). 

\bibitem{Mazroa2009} D. Mazroa, S. Zsigmond, and T. Cinkler, 
Photon. Network Commun. {\bf 18}, 77 (2009).   

\bibitem{Tian2006} B. Colella, F. Effenberger, C. Shimer, 
and F. Tian, ``Raman Crosstalk Control in Passive Optical Networks'', 
in {\it Proc. Fiber Opt. Eng. Conf.}, Anaheim, CA,  
2006, paper NWD6.  

\bibitem{Kim2007} H. Kim, S. B. Jun, and Y. C. Chung, 
IEEE Photon. Technol. Lett. {\bf 19}, 695 (2007).  

\bibitem{Mokhtar2009} Z. A. T. Al-Qazwini, M. K. Abdullah, 
and M. B. Mokhtar, Opt. Eng. {\bf 48}, 015001 (2009). 

\bibitem{Vanholsbeeck2005} F. Vanholsbeeck, S. Coen, P. Emplit, 
M. Haelterman and T. Sylvestre,  Opt. Commun. {\bf 250}, 191 (2005).

\bibitem{Xu2004} C. Xu, X. Liu, and X. Wei, 
IEEE J. Quantum Electron. {\bf 10}, 281 (2004).

\bibitem{Gnauck2005} A. H. Gnauck and P. J. Winzer, 
J. Lightwave Technol. {\bf 23}, 115 (2005). 

\bibitem{Golovchenko2000} V. J. Mazurczyk, G. Shaulov, and 
E. A. Golovchenko, IEEE Photon. Technol. Lett. {\bf 12}, 1573 (2000). 

\bibitem{Kim2008} H. Kim, J. H. Lee, and H. Ji, Opt. Express {\bf 16}, 
20687 (2008). 

\bibitem{Zakharov72} V. E. Zakharov and A. B. Shabat,   
Sov. Phys. JETP {\bf 34}, 62 (1972).


\bibitem{Islam2004} M. N. Islam, ed.,  
{\it Raman Amplifiers for Telecommunications 1: Physical Principles}  
(Springer, New York, 2004).

\bibitem{Agrawal2005} C. Headley and G. P. Agrawal, eds., 
{\it Raman Amplification in Fiber Optical Communication Systems} 
(Elsevier, San Diego, CA, 2005).

\bibitem{Mollenauer2003} L. F. Mollenauer, A. Grant, X. Liu, 
X. Wei, C. Xie, and I. Kang, Opt. Lett. {\bf 28}, 2043 (2003). 

\bibitem{Grosz2004} D. F. Grosz, A. Agarwal, S. Banerjee, D. N. Maywar, 
A. P. K\"ung, J. Lightwave Technol. {\bf 22}, 423 (2004). 


\bibitem{Mamyshev2004} C. Rasmussen, T. Fjelde, J. Bennike, F. Liu, S. Dey, 
B. Mikkelsen, P. Mamyshev, P. Serbe, P. van der Wagt, Y. Akasaka, 
D. Harris, D. Gapontsev, V. Ivshin, and P. Reeves-Hall, 
J. Lightwave Technol. {\bf 22}, 203 (2004).


\bibitem{Rottwitt2004} Z. B. Xu, K. Rottwitt, C. Peucheret, 
and P. Jeppesen, IEEE Photon. Technol. Lett. {\bf 16}, 329 (2004). 

\bibitem{Turitsyn2006}  J. D. Ania-Casta\~n\'on, T. J. Ellingham, 
R. Ibbotson, X. Chen, L. Zhang, and S. K. Turitsyn, 
Phys. Rev. Lett. {\bf 96}, 023902 (2006). 



\bibitem{MM98} L. F. Mollenauer and P. V. Mamyshev,
IEEE J. Quantum Electron. {\bf 34}, 2089 (1998).


\bibitem{Gnauck2008} A. H. Gnauck, R. W. Tkach, A. R. Chraplyvy, 
and T. Li, J. Lightwave Technol. {\bf 26}, 1032 (2008).  



\bibitem{Volterra26} V. Volterra, ``Variations and fluctuations of the 
number of individuals in animal species living together'', translated in 
{\it Animal Ecology}, R. N. Chapman, ed., (McGraw-Hill, 1931, New York).


\bibitem{Stolen89} R. H. Stolen, J. P. Gordon, W. J. Tomlinson, 
and H. A. Haus, J. Opt. Soc. Am. B {\bf 6}, 1159 (1989).


\bibitem{dimensions} The dimensionless $z$ in Eq. (\ref{cfs1}) is
$z=(|\beta_{2}|X)/(2\tau_{0}^{2})$, where $X$ is the actual position,
$\tau_{0}$ is the soliton width, and $\beta_{2}$ is the second order
dispersion coefficient. The dimensionless retarded time is
$t=\tau/\tau_{0}$, where $\tau$ is the retarded time.  
The spectral width is $\nu_{0}=1/(\pi^{2}\tau_{0})$ and the 
frequency difference is $\Delta\nu=(\pi\Delta\beta\nu_{0})/2$. 
$\psi=E/\sqrt{P_{0}}$, where $E$ is proportional to the 
electric field and $P_{0}$ is the peak power. 
The dimensionless second order dispersion coefficient is 
$d=-1=\beta_{2}/(\gamma P_{0}\tau_{0}^{2})$, 
where $\gamma$ is the Kerr nonlinearity coefficient.
The coefficient $\epsilon_{R}$ is given by 
$\epsilon_{R}=0.006/\tau_{0}$, where $\tau_{0}$ 
is in picoseconds.



\bibitem{Malomed91b} B. A. Malomed, Phys. Rev. A {\bf 43}, 3114  (1991).

\bibitem{PCG2003} A. Peleg, M. Chertkov, and I. Gabitov, 
Phys. Rev. E {\bf 68}, 026605 (2003).

\bibitem{PCG2004} A. Peleg, M. Chertkov, and I. Gabitov, 
J. Opt. Soc. Am. B {\bf 21}, 18 (2004).


\bibitem{OOK} 
It is possible to show that the same model describes the 
drift part of the amplitude dynamics in RZ OOK transmission systems. 
See Refs. \cite{P2007,CP2008} for a detailed derivation. 
Note that in this case the coefficient $C$ 
in Eq. (\ref{crosstalk5}) is given by $C=4\epsilon_{R}s\Delta\beta/T$, 
where $s$ is the average fraction of occupied time slots.    


\bibitem{2N} For transmission with an even number of channels 
Eq. (\ref{crosstalk7a}) does not necessarily possess solutions 
other than the trivial solution. However, when the Raman gain 
is described by the triangular approximation Eq. (\ref{crosstalk7a}) 
has infinitely many solutions even for a $2N$-channel system. 
These solutions are given by expressions similar to the ones in 
Eq. (\ref{crosstalk7b}). The proof of stability of the corresponding 
equilibrium states is similar to the proof for transmission with an 
odd number of channels. 


\bibitem{Smale74} M. W. Hirsch and S. Smale, 
{\it Differential Equations, Dynamical Systems, and Linear Algebra}
(Academic Press, New York, 1974).

\bibitem{Wiggins91} S. Wiggins, {\it Introduction to Applied Nonlinear 
Dynamical Systems and Chaos} (Springer, New York, 2003). 


\bibitem{Z_p} It is possible to show that $Z_{p}$ is inversely 
proportional to $\epsilon_{R}$. 


\bibitem{Cuenot2003} B. Cuenot, IEEE Photon. Technol. Lett. {\bf 15}, 
864 (2003).   


\bibitem{Gnauck2003} A. H. Gnauck, G. Raybon, P. G. Bernasconi, 
J. Leuthold, C. R. Doerr, and L. W. Stulz, 
IEEE Photon. Technol. Lett. {\bf 15}, 1618 (2003).


\bibitem{Morita2006} M. Daikoku, T. Miyazaki, I. Morita, H. Tanaka, 
F. Kubota, and M. Suzuki, IEEE Photon. Technol. Lett. {\bf 18}, 391 (2006).  
 

\bibitem{Sandel2007} A. F. Abas, A. Hidayat, D. Sandel, 
B. Milivojevic, and R. Noe, Opt. Fiber Technol. {\bf 13}, 46 (2007).


\bibitem{Mitschke86} F. M. Mitschke and L. F. Mollenauer, 
Opt. Lett. {\bf 11}, 659 (1986).


\bibitem{Gordon86} J. P. Gordon, Opt. Lett. {\bf 11}, 662 (1986).


\bibitem{Kodama87} Y. Kodama and A. Hasegawa, 
IEEE J. Quantum Electron. {\bf 23}, 510 (1987).

\bibitem{Agrawal96} C. Headley III and G. P. Agrawal, 
J. Opt. Soc. Am. B {\bf 13}, 2170 (1996).

\end{thebibliography}
\end{document}